\documentclass[apj]{emulateapj}
\usepackage{apjfonts}
\usepackage{color}

\DeclareMathAlphabet{\mathbi}{OT1}{ptm}{bx}{it}
\SetMathAlphabet\mathbi{bold}{OT1}{ptm}{bx}{it}

\slugcomment{To appear in {\it The Astrophysical Journal.}}
\lefthead{LI et al.}
\righthead{}
\begin{document}
\title{Alignments Of Black Holes With Their Warped Accretion Disks And Episodic Lifetimes
Of Active Galactic Nuclei}
\author{
  Yan-Rong Li\altaffilmark{1}, Jian-Min Wang\altaffilmark{1,2} , Cheng Cheng\altaffilmark{2} ,
  and Jie Qiu\altaffilmark{1}
  }
\altaffiltext{1}
{
Key Laboratory for Particle Astrophysics, Institute of High
Energy Physics, Chinese Academy of Sciences, 19B Yuquan Road,
Beijing 100049, China; liyanrong@mail.ihep.ac.cn
}

\altaffiltext{2}
{
National Astronomical Observatories of China, Chinese
Academy of Sciences, 20A Datun Road, Beijing 100020, China
}

\begin{abstract}
Warped accretion disks have attracted intensive attention because of their critical role on
shaping the spin of supermassive massive black holes (SMBHs) through the Bardeen-Petterson
effect, a general relativistic effect that leads to final alignments or anti-alignments between
black holes and warped accretion disks. We study such alignment processes by
explicitly taking into account the finite sizes of accretion disks and the episodic lifetimes
of AGNs that delineate the duration of gas fueling onto accretion disks. We employ an
approximate global model to simulate the evolution of accretion disks, allowing to determine
the gravitomagnetic torque that drives the alignments in a quite simple way.
We then track down the evolutionary paths for mass and spin of black holes both in
a single activity episode and over a series of episodes. Given with randomly and isotropically
oriented gas fueling over episodes, we calculate the spin evolution with different episodic
lifetimes and find that it is quite sensitive to the lifetimes. We therefore propose that
spin distribution of SMBHs can place constraints on the episodic lifetimes of AGNs and vice versa.
Applications of our results on the observed spin distributions of SMBHs and the observed episodic
lifetimes of AGNs are discussed, although both the measurements at present are yet ambiguous to
draw a firm conclusion.  Our prescription can be easily incorporated into semi-analytic models
for black hole growth and spin evolution.
\end{abstract}
\keywords{accretion, accretion disks --- black hole physics --- galaxies: active --- quasars: general}

\section{Introduction}
Accretion disks surrounding supermassive black holes (SMBHs) are believed to
be the source engines of enormous radiative power of active galactic nuclei (AGNs)
in the universe (e.g., \citealt{Salpeter1964, Lynden1969}). The shapes of accretion
disks---flat or warped---depend on the orientation of the angular momentum of gas fueling from
the circumnuclear region with respect to the rotating orientation of the central black hole.
A wide variety of evidence have indicated that in many circumstances these two orientations
are misaligned.

First, the orientation of AGNs seems uncorrelated with that of their host galaxies,
well established by observations (\citealt{Kinney2000, Schmitt2003, Gallimore2006,
Munoz2007, Shen2010, Lagos2011}) and also confirmed by numerical simulations
(e.g., \citealt{Hopkins2012} and references therein). This suggests that
gas channeled onto the central SMBH either loses the memory of its initial direction
in the host galaxy or comes from outside the galaxy (e.g., \citealt{Dubois2012, Nayakshin2012,
Gaspari2013}), most likely due to galaxy mergers (\citealt{Kendall2003}).
Secondly, the current hierarchical framework of galaxy formation and evolution
predicts that repeat galaxy mergers, in addition to secular evolution,  trigger SMBH
activities (e.g., \citealt{Benson2010}) and gas accretion plausibly proceeds at episodic
and random phases (in particular for minor mergers; e.g., \citealt{King2006, Wang2006}).
Thirdly, recent studies on SMBH spin through quantifying the radiative efficiency of
AGN populations under the thin accretion disk model showed that SMBH spin
undergoes a decline with cosmic time sine $z\sim2$ (\citealt{Wang2009, Li2012}),
potentially implying that (partially) random accretion takes place
(\citealt{King2008}). Such evolutionary behavior has also been  reproduced by
\cite{Volonteri2013} in their semi-analytic model of SMBH growth and evolution
(see their Figure 10).

Once gas fueling is inclined with respect to the central spinning black hole,
a warped accretion disk forms inevitably. The Lense-Thirring torque arising from
the frame dragging effect will lead to precessions of the inclined accretion disk
and the black hole around each other. Detailed dynamics of warped accretion disks
has been studied extensively through theoretical analysis (e.g., \citealt{Pringle1992,
Papaloizou1995, Ogilvie1999, Lubow2002}) and numerical simulations
(e.g., \citealt{Nelson1999, Nelson2000, Lodato2006, Lodato2007,
Fragile2009, Zhuravlev2014}). The intensive applications of warped accretion disks
lie at their critical role on shaping the spin of the central SMBHs through
the Bardeen-Petterson effect (\citealt{Bardeen1975}). In this effect, the presence of
strong viscosity, a most like case for normal AGNs, damps out the differential
Lense-Thirring precession and induce the inner portion of the disk to align
or anti-align with the spin axis of the hole. Accompanied with warp propagation,
the entire disk will tend to align or anti-align with the hole (\citealt{Scheuer1996,
King2005, Martin2007}).

Previous studies on the alignments between black holes and their warped accretion disks
mainly focus on two aspects. The first assumes that accretion disks have infinite
extensions and constant profiles. Black holes are always driven toward
the disks and the black hole-disk systems end up alignments (\citealt{Scheuer1996,
Martin2007, Perego2009}). On the contrary, the second assumes that accretion disks are
finite in both size and mass and the total angular momentum of the systems is conserved
(\citealt{King2005, King2008}), which means that there is no gas supply to the disks. Such systems
can end up a configuration of either alignment or anti-alignment, depending on
the initial inclinations and the ratios of the angular momenta between disks and black
holes (\citealt{King2005}).

In reality, it is well known that accretion disks are truncated by their own self-gravity
in the outer regions where it becomes dominated over the gravity from black holes
(e.g., \citealt{Paczynski1978, Shlosman1987, Collin1999}). Outside the
truncation radius, the disks fragment into dense clumps and finally turn to be a hotbed of
star formations (e.g., \citealt{Goodman2003, Levin2007, Wang2010, Jiang2011}).
It comes naturally that the stellar radiation and supernova explosions feedback to
the ambient gas, exciting turbulence that acts to transfer the angular momentum and
leads to gas supply to the accretion disks from the further outer regions
(\citealt{Wada2002, Kawakatu2008, Kumar2010, Wang2010}). Therefore,
to appropriately study warped accretion disks, it is very practical to
take into account both the finite sizes limited by the self-gravity and the plausible gas
feeding at the outer boundaries (see also the discussions of \citealt{Li2013a}).
The lifetime that the gas feeding lasts for, in principle, can be determined from the
observed episodic lifetime of AGNs, for which there are several techniques,
although the results are still primitive (see \citealt{Martini2004} for a review).

In this paper, we revisit the alignment processes of black holes with their warped accretion
disks by including the finite sizes of accretion disks and the continuous gas feeding
controlled by the episodic lifetimes of AGNs. For the convenience of
numerical implementation, we introduce  a parameter ``the lifetime of gas fueling''
to describe the duration of gas supply. We employ an approximate
global model following \cite{Kumar2008} to describe the accretion process.
This enables us to study the alignments of finite-size accretion disks
in a quite simply way, avoiding being immersed in
the time-consuming numerical simulations on the evolution of accretion disks.

The paper is organized as follows. Section 2 describes the properties of
warped accretion disks and Section 3 derives the basic evolution equations
for the black hole-accretion disk system. In Section 4, we present the results
for alignments/anti-alignments and spin evolution in a single
accretion event and over a series of events. We then show the implications
of our calculations for spin distribution and episodic lifetimes of AGNs
in Section 5. Discussions on plausible caveats of our calculations and conclusions are given in Section~6.

\section{Properties of Warped Accretion Disks}
In the weak-field limit, the frame-dragging effect due to
a spinning black hole exerts a Lense-Thirring torque on
each annulus of the accretion disk (\citealt{Bardeen1975}).
By denoting the angular momentum of the annulus at a radius $R$
as $\mathbi{L}(R)$ and that of the black hole as $\mathbi{J}_{\rm h}$,
the Lense-Thirring torque reads
\begin{equation}
 \mathbi{t}_{\rm LT}=\frac{2G}{c^2}\frac{\mathbi{J}_{\rm h}\times \mathbi{L}(R)}{R^3},
\end{equation}
where $G$ is the gravitational constant and $c$ is the speed of light.
This Lense-Thirring torque leads the annuli to precess around the rotating
axis of the black hole at a rate of $2GJ_{\rm h}/c^2R^3$.
Since such differential precessing rate rapidly declines with radius as $R^{-3}$,
the inner region of the accretion disk is intensely affected in the beginning
and gradually warps propagate outwards over the entire disk.
Detailed theoretical analysis showed that the dynamics of warped accretion disks
is characterized into two regimes depending on the importance of the pressure
force and viscous force (\citealt{Papaloizou1995}). If adopting the standard $\alpha$-prescription of
viscosity, warps propagate as bending waves for inviscid or sufficiently thick
disks ($\alpha<H/R$), whereas propagate in a diffusive fashion for strongly viscous disks
($\alpha>H/R$). Here $H$ is the semi-thickness of disks. For simplicity, we only consider
thin and viscous accretion disks with {\em Keplerian} rotation,
in which $\alpha>H/R$ is generally satisfied.
This is the most likely case for accretion disks in AGNs with the Eddington ratio
at around 0.01-1, the largest population in AGN surveys (e.g., \citealt{McLure2004, Shen2008}).

In the presence of high viscosity, the differential precess in the
inner disk will be damped out rapidly, giving rising to an inner flat region
(so called Bardeen-Petterson effect; \citealt{Bardeen1975}). This region
extends to the radius where the precessing timescale is comparable with
the warp-diffusion timescale, The precessing time scale
reads $c^2R^3/2GJ_{\rm h}$ and the warp-diffusion time scale reads $R^2/\nu_2$,
leading to the warp radius as
\begin{equation}
R_{\rm w} \sim \frac{2GJ_{\rm h}}{\nu_2 c^2},
\label{eqn_warp}
\end{equation}
where $\nu_2$ is the vertical shear viscosity governing the warp diffusion
through the accretion disks.

Integrating over all the annuli yields the total torque exerted on the accretion disk
\begin{equation}
\mathbi{T}_{\rm LT} = \frac{4\pi G}{c^2}\mathbi{J}_{\rm h}\times\int\frac{\mathbi{L}(R)}{R^2}dR.
\label{eqn_torque}
\end{equation}
As a reaction, the black hole, in the meantime, suffers an equal but opposite torque
that cause its angular momentum to process as well. By taking into account the angular momentum
carried by mass accretion, the angular momentum of the black hole evolves following
\begin{equation}
\frac{d\mathbi{J}_{\rm h}}{dt} = \dot M_{\rm h} \ell_{\rm ms} {\hat\mathbi{J}}_{\rm h}-\mathbi{T}_{\rm LT},
\label{eqn_jh}
\end{equation}
where $\dot M_{\rm h}$ is the mass accretion rate onto the black hole and $\ell_{\rm ms}$
is the specific angular momentum at the marginal stable orbit $R_{\rm ms}$. Hereafter
a hat symbol on top of a vector represents the corresponding direction vector.

Similarly, the angular momentum of disks changes with time as
\begin{equation}
\frac{d\mathbi{J}_{\rm d}}{dt} = -\dot M_{\rm h} \ell_{\rm ms} {\hat\mathbi{J}}_{\rm h} +\mathbi{T}_{\rm LT}
+{\dot\mathbi{J}}_{\rm f},
\label{eqn_jd}
\end{equation}
where $\mathbi{J}_{\rm d}$ is angular momentum summed over all
the annuli and ${\dot\mathbi{J}}_{\rm f}$ is angular momentum change due to gas fueling.
Note that $\mathbi{T}_{\rm LT}$ is always orthogonal to $\mathbi{J}_{\rm h}$, therefore, it
can be expressed in a form of (\citealt{King2005})
\begin{equation}
\mathbi{T}_{\rm LT} = K_1\hat{\mathbi{J}}_{\rm h}\times\hat{\mathbi{J}}_{\rm d}
+K_2\hat{\mathbi{J}}_{\rm h}\times(\hat{\mathbi{J}}_{\rm h}\times\hat{\mathbi{J}}_{\rm d}),
\end{equation}
where the first term in the right hand side corresponds to the precession and
the second term corresponds to change in the angle between the hole and the disk.
We keep in mind that both $K_1$ and $K_2$ are generally positive as shown in our previous
numerical analysis (\citealt{Li2013a}).

Considering that the inner region of the disk is aligned with the black hole and
the Lense-Thirring torque drops off rapidly with radius, the major
contribution to $\mathbi{T}_{\rm LT}$ in Equation (\ref{eqn_torque}) comes from
the integral around the warp radius $R_{\rm w}$. We therefore make an approximate
for $K_2$ as
\begin{equation}
K_2\approx\frac{2\pi G}{c^2}\frac{J_{\rm h}L(R_{\rm w})}{R_{\rm w}}.
\label{eqn_k2}
\end{equation}
Below we will show that such an approximation of $K_2$ yields results that
exactly agree with the previous analytical studies for small warps
(\citealt{Scheuer1996, Martin2007}). Note that $K_1$ only governs the precession
of the black hole-disk system and therefore is not important for the present
purpose.

\begin{figure*}[ht!]
\centering
\includegraphics[angle=-90.0, width=0.4\textwidth]{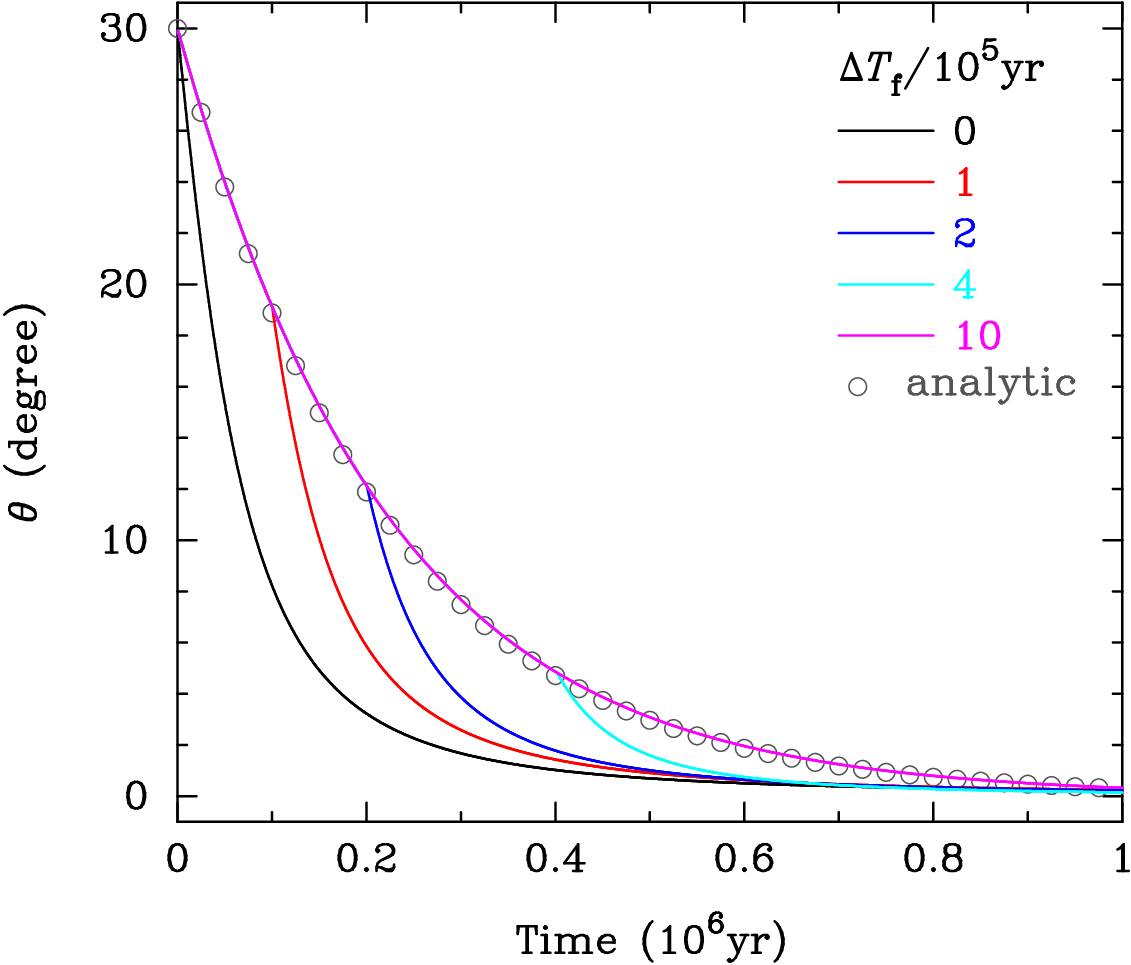}~~~~~~~
\includegraphics[angle=-90.0, width=0.4\textwidth]{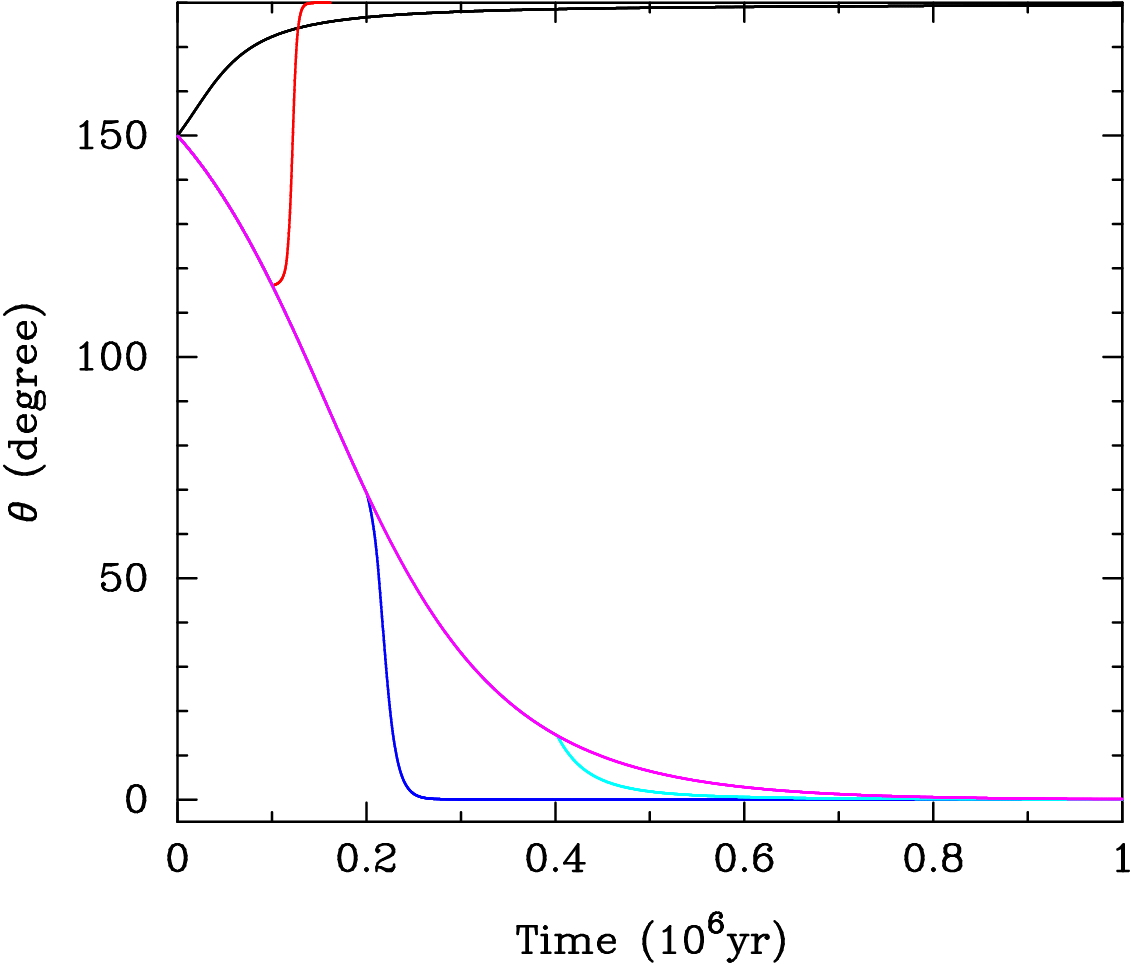}
\caption{Evolution of inclination angle $\theta$ between black holes and disks for different lifetimes of gas fueling
$\Delta T_{\rm f}$.
The initial values are (Left) $\theta=30^\circ$ and (Right) $\theta=150^\circ$, respectively.
Open circles in left panel represent the analytic solution for small warps: an exponentially decay
with an e-folding timescale $T=2.2\times10^5$yr (see Equation (\ref{eqn_angle_anal})).}
\label{fig_angle}
\end{figure*}

\begin{figure*}[ht!]
\centering
\includegraphics[angle=-90.0, width=0.4\textwidth]{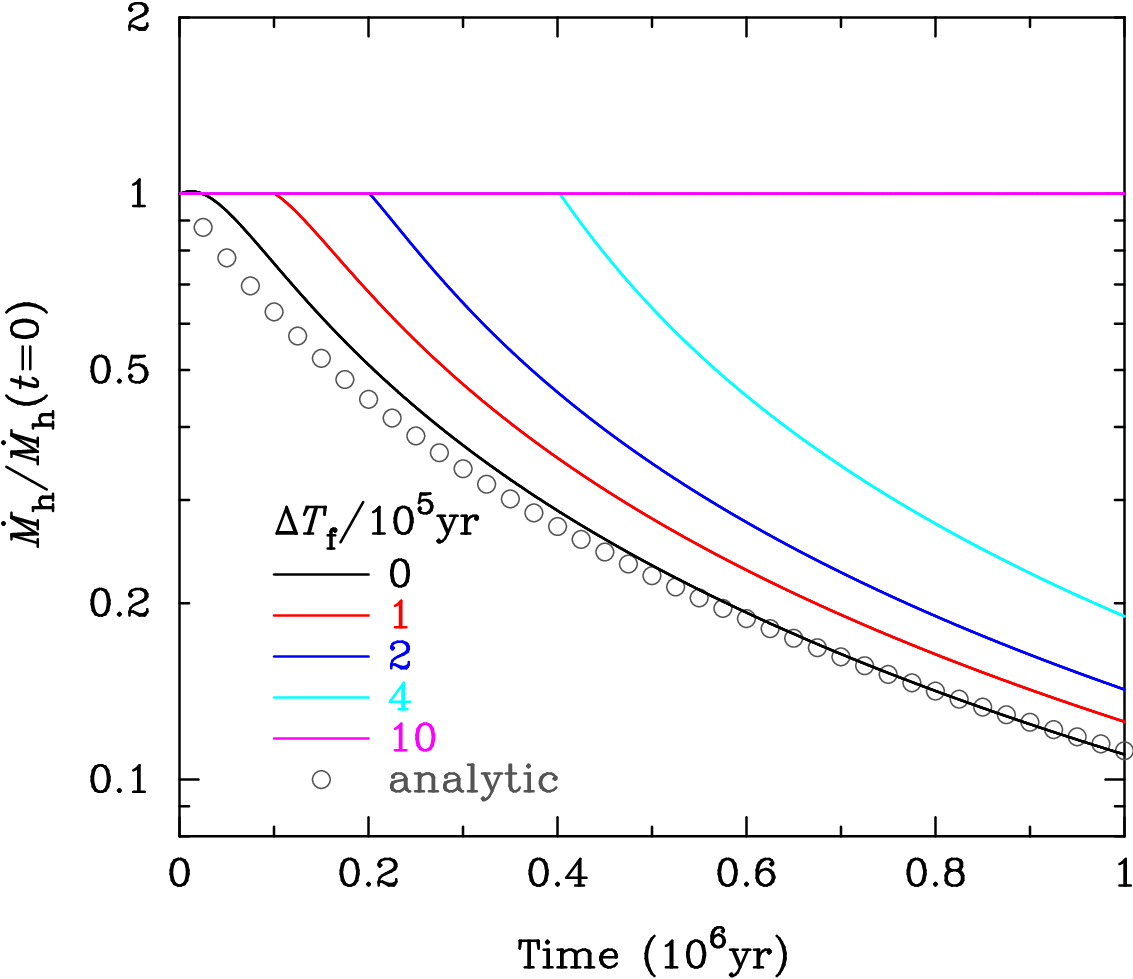}~~~~~~~
\includegraphics[angle=-90.0, width=0.4\textwidth]{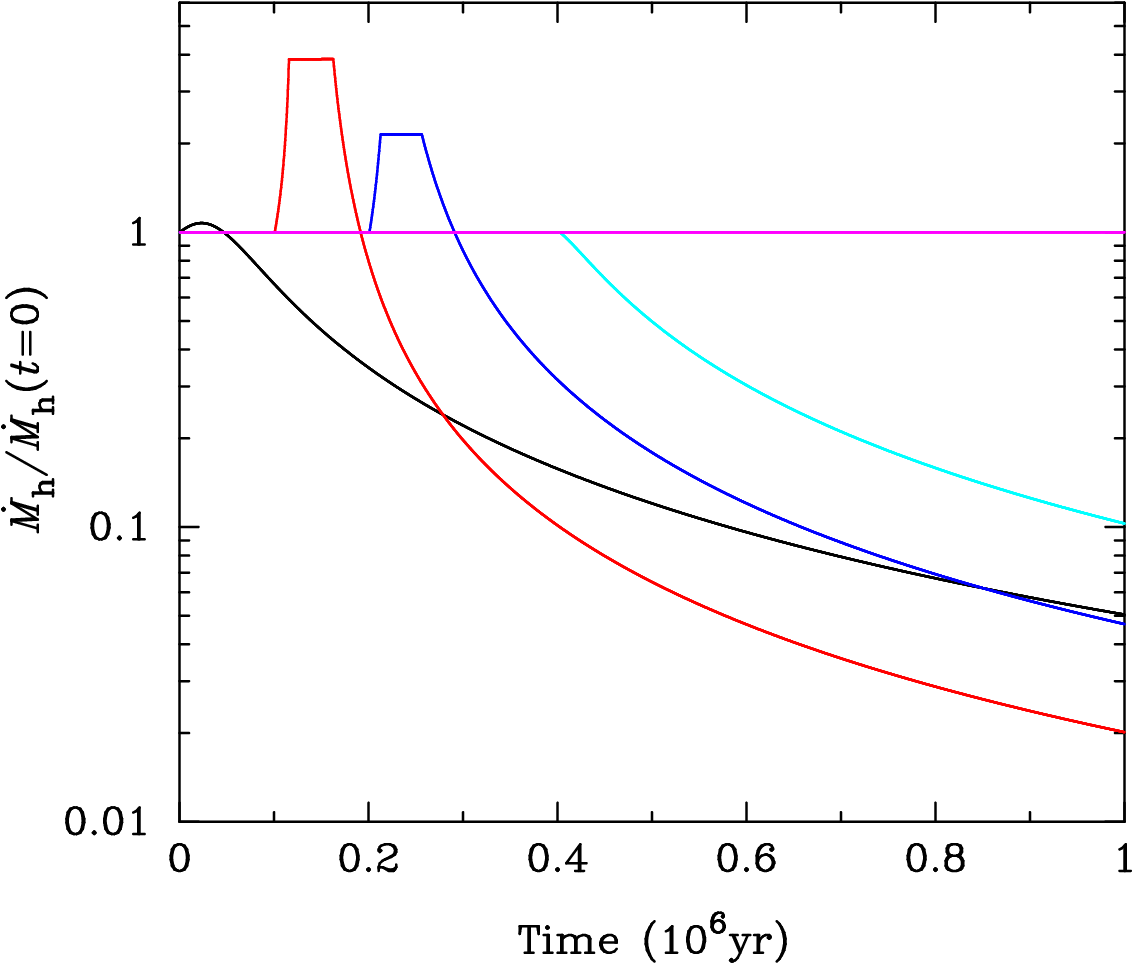}
\caption{Same as Figure \ref{fig_angle} but for mass accretion rate.
The initial mass accretion rate is $\dot M_{\rm h}(t=0)=0.25\dot M_{\rm Edd}$.
In left panel, open circles represent the asymptotic evolution following Equation (\ref{eqn_mdot_asym}).
In right panel, mass accretion rates are limited by the Eddington rate.}
\label{fig_rate}
\end{figure*}

\section{Basic Evolution Equations}
\subsection{Accretion Disks and the Inclination Angles}
From above equations, we can now determine the evolution of the angular momenta
$J_{\rm h}$ and $J_{\rm d}$ and the angle $\theta$ between the black hole and
the accretion disk. We first define a lifetime of gas fueling $\Delta T_{\rm f}$ as
during which the accretion disk is continuously fed at a mass rate limited by the Eddington
rate $\dot M_{\rm Edd}=L_{\rm Edd}/\eta c^2=2.2M_8\eta_{-1}^{-1}M_\odot~{\rm yr}^{-1}$,
where $L_{\rm Edd}$ is the Eddington luminosity, $M_8=M_{\rm h}/10^8M_\odot$, and
$\eta=0.1\eta_{-1}$ is the radiative efficiency for mass accretion
(see below Section~\ref{sec_bhevo} for definition).

Note again the inner region of the disk is aligned with the black hole.
Multiplying ${\hat\mathbi{J}}_{\rm h}$ to Equation~(\ref{eqn_jh}) yields
\begin{equation}
\frac{dJ_{\rm h}}{dt} = \dot M_{\rm h} \ell_{\rm ms},
\label{eqn_jhs}
\end{equation}
and multiplying ${\hat\mathbi{J}}_{\rm d}$ to Equation (\ref{eqn_jd}) yields
\begin{equation}
\frac{dJ_{\rm d}}{dt}=-\dot M_{\rm h} \ell_{\rm ms} \cos\theta - K_2\sin^2\theta + {\dot J}_{\rm f}\cos\theta,
\label{eqn_jds}
\end{equation}
where we assume that the angular momentum of gas fueling is aligned with $\mathbi{J}_{\rm d}$.
The angle between the black hole and the accretion disk changes with time as
\begin{equation}
J_{\rm d}J_{\rm h}\frac{d\cos\theta}{dt} =
\frac{d\mathbi{J}_{\rm h}\cdot\mathbi{J}_{\rm d}}{dt} - J_{\rm h}\cos\theta\frac{dJ_{\rm d}}{dt}
-J_{\rm d}\cos\theta\frac{dJ_{\rm h}}{dt}.
\label{eqn_angle}
\end{equation}
We distinguish the evolution of $\theta$ into two phases as below.

\subsubsection{Phase I: Gas Fueling}
During the phase with gas fueling, the accretion disk is continuously
fed so that it maintains a preferred steady angular momentum distribution.
As a result, one shall expect that
the term due to gas fueling cancels out the
other terms in Equations (\ref{eqn_jd}) and (\ref{eqn_jds}) and
$\mathbi{J}_{\rm d}$ and its magnitude $J_{\rm d}$
remain somewhat constant, i.e., $d\mathbi{J}_{\rm d}/dt=0$ and  $dJ_{\rm d}/dt=0$.
With Equations (3) and (7), we can simplify Equation (\ref{eqn_angle}) into 
\begin{equation}
\frac{d\cos\theta}{dt} = \frac{K_2}{J_{\rm h}}\sin^2\theta.
\label{eqn_angle_p1}
\end{equation}
Since $K_2$ is positive, $\theta$ always tends to decrease towards zero regardless of
its initial value, i.e., alignment between the disk and the black hole
(see also \citealt{Scheuer1996, Martin2007, Perego2009}).

In this phase, as the disk maintains its profile steadily, $K_2$ can be deemed
to be a constant. There exists an analytic solution for Equation (\ref{eqn_angle_p1})
\begin{equation}
\frac{1-\cos\theta}{1+\cos\theta}=\frac{1-\cos\theta_0}{1+\cos\theta_0}\exp\left(-\frac{2t}{T}\right),
\label{eqn_sol_p1}
\end{equation}
where $\theta_0$ is the initial value of $\theta(t)$ at $t=0$ and $T=J_{\rm h}/K_2=
c^2 R_{\rm w}/2\pi G L(R_{\rm w})$.

For small warps, we extend the trigonometric functions
in Equation (\ref{eqn_sol_p1}) around $\theta\sim0$ and it is trivial to
show that the corresponding solution decays exponentially as
\begin{equation}
\theta(t) = \theta_0\exp\left(-\frac{t}{T}\right).
\label{eqn_angle_anal}
\end{equation}
This is the analytic solution found in \citet[see their Equation (11)]{Scheuer1996}
\footnote{Note that \cite{Scheuer1996} omitted a factor of $\sqrt{2}$ in their Equation (11).}
and \citet[see their Equation (52)]{Martin2007} and the time scale of the exponential decay $T$
exactly agrees with their derivations.

\subsubsection{Phase II: Gas Fueling Quenched}
After gas fueling is quenched, i.e., $\dot{J}_{\rm f}=0$, the total
angular momentum of the system (black hole and accretion disk) becomes conserved.
The disk subjects to both alignment and viscous diffusion. The former decreases the
angular momentum of the disk ($K_2>0$ in Equation (\ref{eqn_jds})), whereas
the later transfers the angular momentum outwards.
Substituting Equations (\ref{eqn_jhs}) and (\ref{eqn_jds}) into
Equation (\ref{eqn_angle}) and with some simple mathematical implementations, we arrive at
\begin{equation}
\frac{d\cos\theta}{dt} =-\frac{\dot M_{\rm h} \ell_{\rm ms}}{J_{\rm d}}\sin^2\theta
+ \frac{K_2}{J_{\rm h}}\sin^2\theta\left(1+\frac{J_{\rm h}}{J_{\rm d}}\cos\theta\right).
\label{eqn_angle_p2}
\end{equation}

Following \cite{Kumar2008}, we employ an approximate global model to simulate
the influence of mass accretion on the disk properties so that accretion rate $\dot M$
can be determined in a quite simple fashion.
To this end, we define the effective radius of the disk $R_{\rm d}$ as%
\footnote{We stress that the definition of $R_{\rm d}$ aims to simulate the disk evolution
in a global way. It has the same order of magnitude as, but is not necessarily identical to,
the disk's physical outer truncated radius.}
\begin{equation}
\frac{J_{\rm d}}{M_{\rm d}} = \ell(R_{\rm d})=(GM_{\rm h} R_{\rm d})^{1/2}.
\label{eqn_rd}
\end{equation}
The mean global accretion rate is then given by
\begin{equation}
\dot M_{\rm h} = -\frac{dM_{\rm d}}{dt} = \frac{M_{\rm d}}{t_{\rm acc}},
\label{eqn_mdot1}
\end{equation}
where the accretion time scale reads
\begin{equation}
t_{\rm acc}\sim\frac{R_{\rm d}^2}{\nu_1(R_{\rm d})},
%= \frac{1}{\alpha h^2 \Omega_{\rm K}(R_{\rm d})}
%=\frac{J_{\rm d}^3}{\alpha h^2 G^2 M_{\rm h}^2 M_{\rm d}^3},
\end{equation}
where $\nu_1$ is the normal viscosity associated to the azimuthal shear due to
the differential rotation of the disk. With the standard $\alpha$-prescription,
$\nu_1$ is expressed by
\begin{equation}
\nu_1=\alpha \Omega_{\rm K} H^2=\alpha h^2\Omega_{\rm K} R^2,
\end{equation}
where $h=H/R$ is the disk's aspect ratio and $\Omega_{\rm K}$ is the Keplerian
angular momentum. Combining above equations, we can obtain
\begin{equation}
\dot M_{\rm h} = \frac{\alpha h^2 G^2M_{\rm h}^2 M_{\rm d}^4}{J_{\rm d}^3}.
\label{eqn_mdot2}
\end{equation}
This equation apparently indicates the influences of
alignment and accretion on $\dot M_{\rm h}$. From Equation (\ref{eqn_jds}),
alignment processes reduce the disk angular momentum
 $J_{\rm d}$ at a rate of $\sim K_2$ but leave the disk mass $M_{\rm d}$
unchanged, therefore, increasing the accretion rate $\dot M_{\rm h}$.
On the contrary, mass accretion reduces the disk mass
but approximately keeps the disk angular momentum unchanged, considering
that angular momentum loss due to the accretion inflow through the marginal stable
orbit is usually negligible, therefore decreasing $\dot M_{\rm h}$.

\begin{figure*}[th!]
\centering
\includegraphics[angle=-90.0, width=0.4\textwidth]{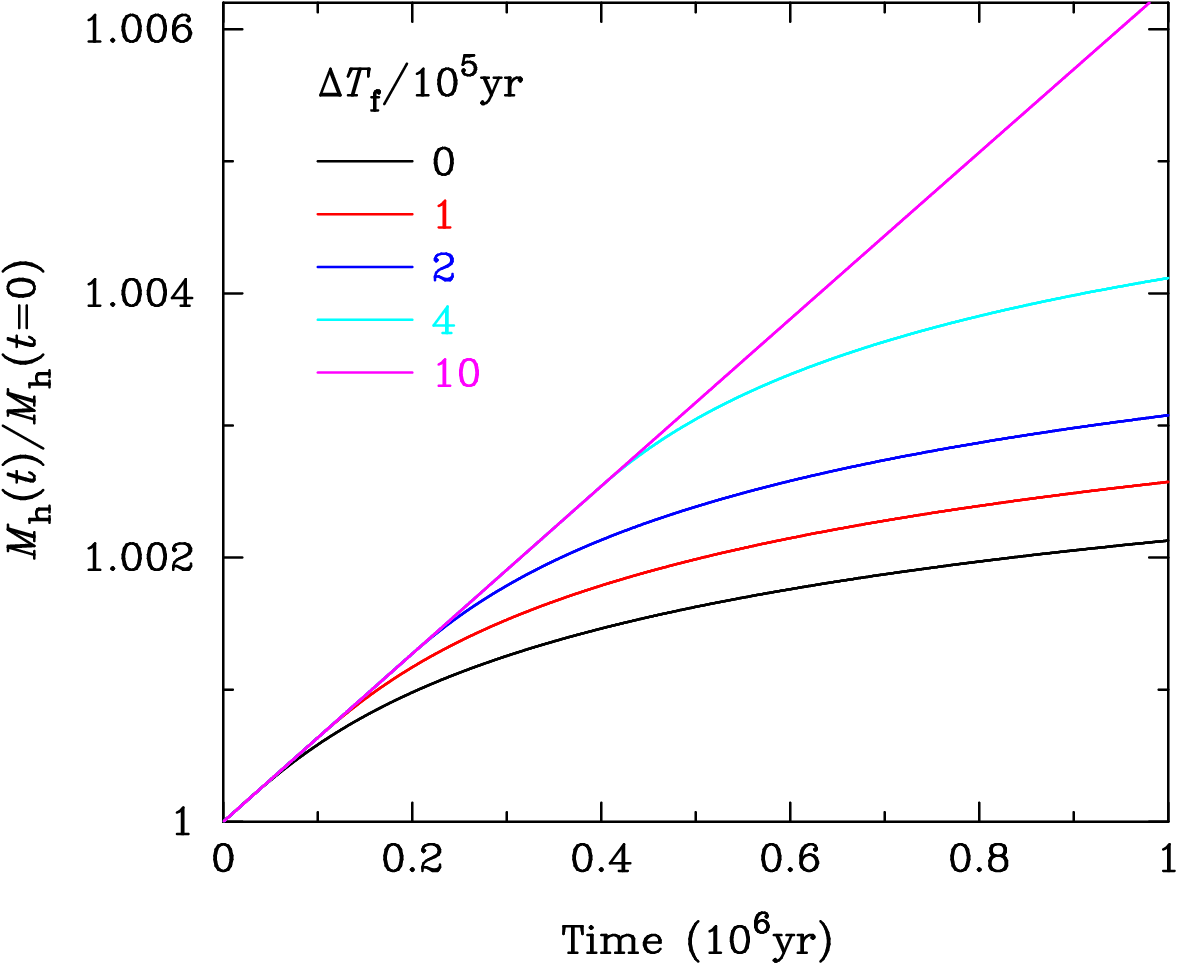}~~~~~~~
\includegraphics[angle=-90.0, width=0.4\textwidth]{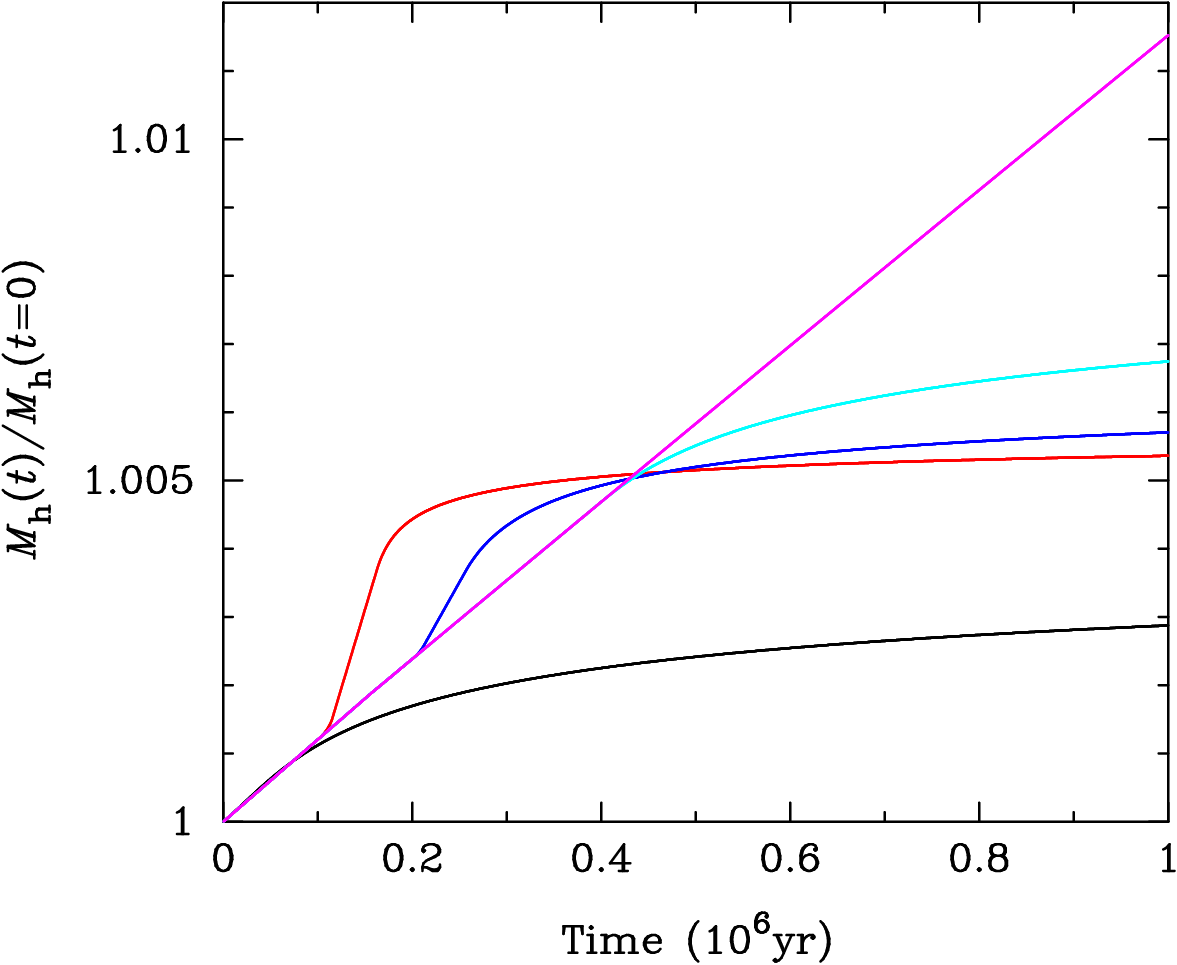}
\caption{Same as Figure \ref{fig_angle} but for mass growth of black holes.
The initial mass of the black hole is $M_{\rm h}(t=0)=10^8M_\odot$.}
\label{fig_mass}
\end{figure*}
\begin{figure*}[th!]
\centering
\includegraphics[angle=-90.0, width=0.4\textwidth]{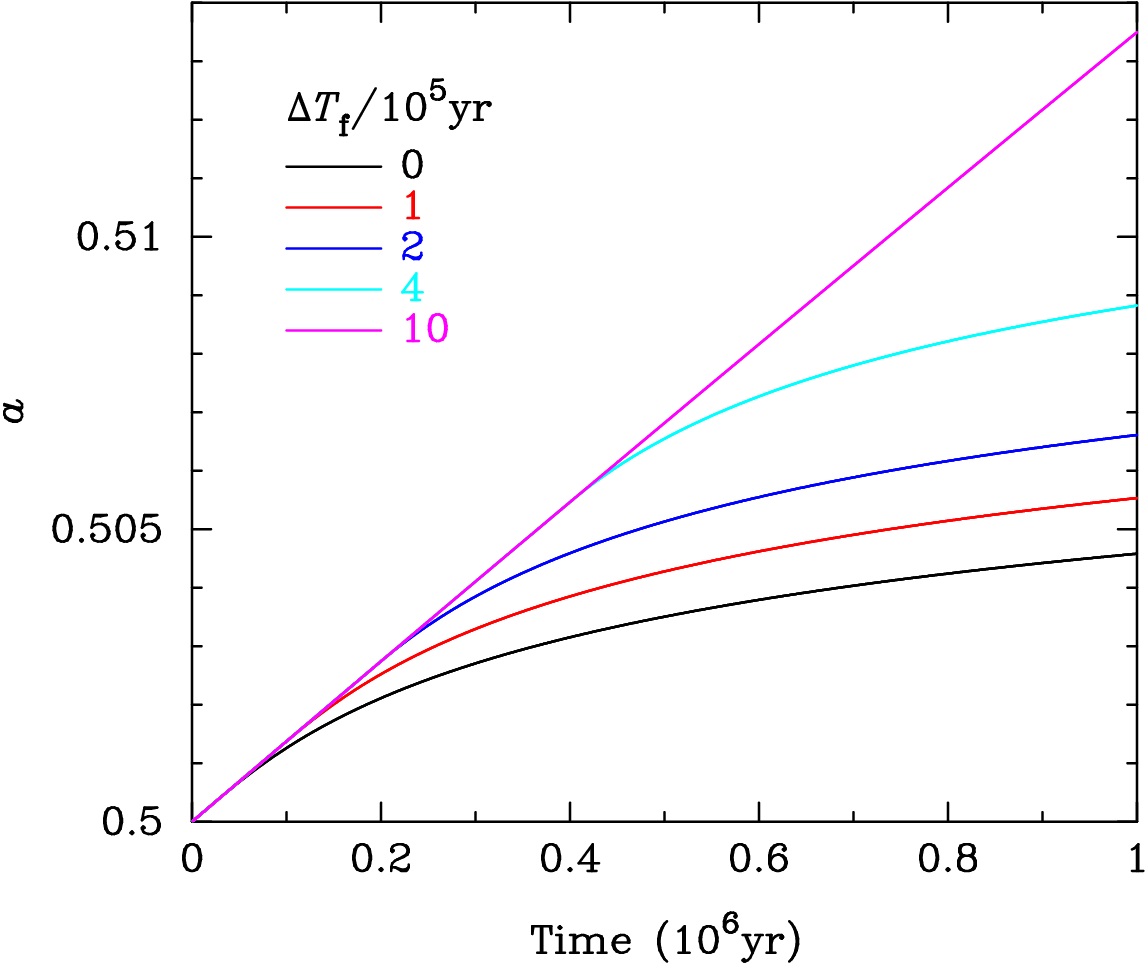}~~~~~~~
\includegraphics[angle=-90.0, width=0.4\textwidth]{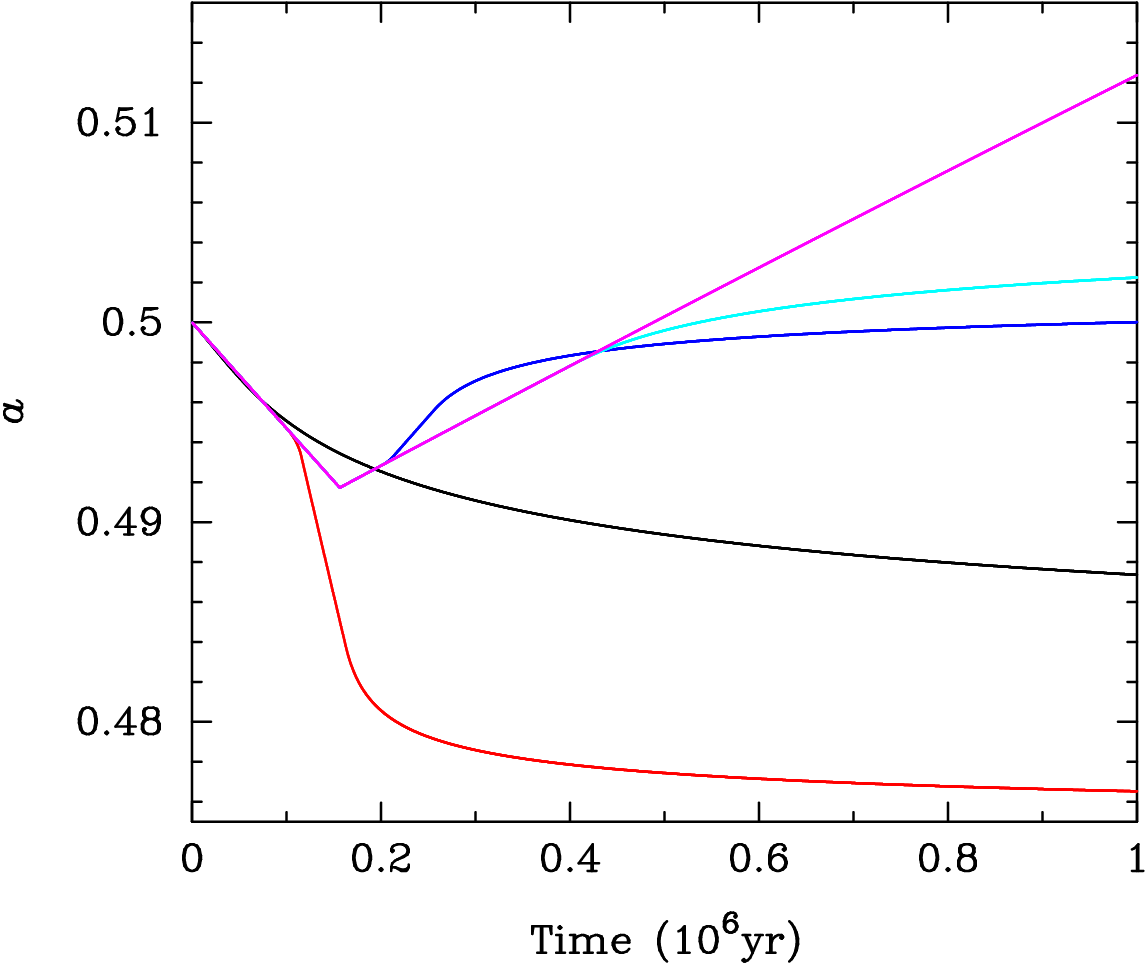}
\caption{Same as Figure \ref{fig_angle} but for spin evolution of black holes.
The initial black hole spin is $a=0.5$. }
\label{fig_spin}
\end{figure*}

Indeed, if only considering the influence of accretion rate,
there exists an asymptotic solution to Equation (\ref{eqn_mdot1})
(see \citealt{Kumar2008})
\begin{equation}
M_{\rm d}(t) = M_{\rm d}(t=0) \left[1 +\frac{3 t}{t_{\rm acc}'} \right]^{-1/3},
\end{equation}
and
\begin{equation}
\dot M_{\rm h}(t) = \frac{M_{\rm d}(t=0)}{t_{\rm acc}'}
\left[1+\frac{3 t}{t_{\rm acc}'}\right]^{-4/3},
\label{eqn_mdot_asym}
\end{equation}
where { $t_{\rm acc}'=t_{\rm acc}(0)$} is the accretion time scale at $t=0$.
For small warp cases, we expect that the mass accretion
rate approaches the above asymptotic behaviors, i.e., $\dot M_{\rm h}\propto t^{-4/3}$
for $t\gg t_{\rm acc}'$.

With mass accretion rate, we now rewrite $K_2$ in a more convenient form.
Previous studies showed that the vertical viscosity $\nu_2$ is related
to the normal azimuthal viscosity $\nu_1$ by (\citealt{Ogilvie1999, Lodato2007};
see also \citealt{Perego2009})
\begin{equation}
\frac{\nu_2}{\nu_1} = \frac{f_\nu}{2\alpha^2},
\label{eqn_vis}
\end{equation}
where $f_\nu$ is a coefficient of order of unity.
As a zeroth-order approximation, we neglect the detailed shape of warped
accretion disks and make use of the estimate for surface density
$\Sigma \approx \dot M_{\rm h}/3\pi\nu_1$ and for the angular momentum
$L(R) \approx \dot M_{\rm h} \sqrt{GM_{\rm h} R}/3\pi\nu_1$ as in flat disks.
This is reasonable since accretion disks reach maximally warped around the
warp radius $R_{\rm w}$ and then rapidly approach flat geometry beyond $R_{\rm w}$
(\citealt{Scheuer1996, Martin2007, Chen2009, Li2013a}).
Combining Equations (\ref{eqn_warp}) and (\ref{eqn_k2}) gives
\begin{equation}
K_2\approx\frac{f_{\nu}\dot M_{\rm h}}{6\alpha^2}\sqrt{GM_{\rm h} R_{\rm w}}.
\label{eqn_K2_mdot}
\end{equation}
For small warp amplitude, $f_\nu=1$, whereas for large warp amplitude,
$f_\nu$ moderately decreases below unity, depending on the warp amplitude 
(\citealt{Ogilvie1999, Lodato2010}). \cite{Ogilvie1999} developed a prescription 
to determine $f_\nu$ for any warp amplitude. However, this needs to know the 
detailed warp profile as a function of radius, therefore beyond the scope of 
the present work. We simply set $f_\nu=1$ throughout our calculations and 
below we will show the reliability of our results in light of this simplification.
%Numerical simulations by \cite{Lodato2006} showed that
%the above equation provides a quite good estimate to the Lense-Thirring torque
%(see their Figure 2).

\subsection{Black Hole Evolution}
\label{sec_bhevo}
The direction of the inner flat region of the disk depends
on the inclinations of the incoming gas flow and the black hole:
if $\theta\leq\pi/2$, it aligns with the hole; while if
$\theta>\pi/2$, it anti-aligns with the hole.
By taking into account the mass loss carried away by radiation, one obtains
that the black hole grows as
\begin{equation}
\frac{dM_{\rm h}}{dt} = (1-\eta)\dot M_{\rm h},
\label{eqn_mh}
\end{equation}
and by combining with Equation (\ref{eqn_jhs}), the black hole spin evolves as
\begin{equation}
\frac{da}{dt} = (1-\eta)\frac{\dot M_{\rm h}}{M_{\rm h}}
\left(\frac{1}{1-\eta}\frac{\ell_{\rm ms}}{R_{\rm g} c} - 2a\right),
\label{eqn_spin}
\end{equation}
where $\eta$ is the radiative efficiency depending on the spin parameter $a$ via
\begin{equation}
\eta = 1 - \sqrt{1-\frac{2}{3r}},
\label{eqn_eta}
\end{equation}
and the specific angular momentum is calculated as
\begin{equation}
\ell_{\rm ms} = \pm\sqrt{GM_{\rm h}R_{\rm ms}}
\frac{1\mp2a r^{-3/2} + a^2 r^{-2}}{(1-3 r^{-1} \pm 2 a r^{-3/2})^{1/2}},
\label{eqn_lms1}
\end{equation}
for $a<1$ and
\begin{equation}
\ell_{\rm ms} = \pm\sqrt{GM_{\rm h}R_{\rm ms}}
\frac{1\pm r^{-1/2} + r^{-1}\mp r^{-3/2}}{(1\pm 2 r^{-1/2})^{1/2}},
\label{eqn_lms2}
\end{equation}
for $a=1$, where $r=R_{\rm ms}/R_{\rm g}$, $R_{\rm g}=GM_{\rm h}/c^2$ is the gravitational radius,
and the upper sign  refers to prograde accretion while the lower sign refers to retrograde
accretion (\citealt{Bardeen1972})
\footnote{Note that, according to the calculation of \cite{Thorne1974},
the radiation emitted by the geometrically-thin disk and swallowed by the hole produces
a counteracting torque that prevents the hole span up beyond an upper limit of $a\approx0.998$.
In our calculations, the accretion may proceed from retrograde to prograde fashion over episodes.
This effectively means than the black hole spin may suffer a flip from $a=-1$ to $a=1$.
For the sake of simplicity, we neglect the counteracting torque and assume that
black holes can carry any spin between -1 and 1.}.

\section{Results}
\label{sec_results}
According to the standard accretion disk model, the disk's aspect ratio $h=H/R$ is in an order of
$10^{-3}\sim10^{-2}$ and weakly depends on radius in proportional to $R^{-1/20}$,
on black hole mass in proportional to $M_{\rm h}^{-1/10}$, and on mass accretion rate
in proportional to $\dot M_{\rm h}^{1/5}$
(\citealt{Shakura1973}; see also \citealt{Collin1990, Natarajan1998, King2008}).
An accretion disk becomes self-gravitating when its mass reaches a factor $h$ of the black hole mass
(e.g., \citealt{Pringle1981}). We therefore adopt the initial disk mass $M_{\rm d}\approx hM_{\rm h}$.
The initial effective radius of the disk $R_{\rm d}$ is determined
by the required {\it initial} mass accretion rate and accretion time scale
given with $h$. To be specific, from Equation (\ref{eqn_mdot2}),
$\dot M_{\rm h}$ can be alternatively expressed through $R_{\rm d}$
and $h$ by
\begin{eqnarray}
\frac{\dot M_{\rm h}}{\dot M_{\rm Edd}} &=& 0.25 \left(\frac{\alpha}{0.1}\right)\left(\frac{\eta}{0.1}\right)
\left(\frac{h}{5\times10^{-3}}\right)^3 \nonumber\\
&&\times\left(\frac{R_{\rm d}}{2.7\times10^3R_{\rm g}}\right)^{-3/2}
\left(\frac{M_{\rm h}}{10^8M_\odot}\right)^{-1}.
\label{eqn_mdot_edd}
\end{eqnarray}
The corresponding accretion time scale is
\begin{eqnarray}
t_{\rm acc}&=&8.8\times 10^5{\rm yr}\left(\frac{\alpha}{0.1}\right)^{-1}
\left(\frac{h}{5\times10^{-3}}\right)^{-2}\nonumber\\
&&\times\left(\frac{R_{\rm d}}{2.7\times10^3R_{\rm g}}\right)^{3/2}
\left(\frac{M_{\rm h}}{10^8M_\odot}\right).
\label{eqn_tacc}
\end{eqnarray}
This is the typical timescale for accretion disks truncated by self-gravitational collapse
(see, e.g., \citealt{Perego2009}). The ratio between the angular momenta of the disk
and the hole is
\begin{equation}
\frac{J_{\rm d}}{J_{\rm h}} = 0.26a^{-1}\left(\frac{h}{5\times10^{-3}}\right)
\left(\frac{R_{\rm d}}{2.7\times10^3R_{\rm g}}\right)^{1/2}.
\label{eqn_ratio}
\end{equation}

Our calculating procedure is as follow. We start with a black hole that 
has a mass of $M_{\rm h}$ and a spin of $a$. The accretion disk is assigned 
an initial mass of $M_{\rm d}=hM_{\rm h}$ and angular momentum of 
$J_{\rm d}=M_{\rm d}(GM_{\rm h}R_{\rm d})^{1/2}$.
The initial inclination angle between the black hole hole and the disk is~$\theta$.
We then evolve Equations (\ref{eqn_jds}) and (\ref{eqn_mdot1}) for
the angular momentum and mass of the disk,  Equations (\ref{eqn_mh})
and (\ref{eqn_spin}) for the mass and spin of the black hole, and finally
Equations (\ref{eqn_angle_p1}) and (\ref{eqn_angle_p2}) for the inclination
angle. The mass accretion rate $\dot M_{\rm h}$ is determined by Equation (\ref{eqn_mdot2}).
We keep in mind that in the phase with gas fueling, the disk's mass and 
angular momentum are unchanged. For a series of accretion events, we just 
repeat the above procedure but with the mass and spin of
the black hole inherit from the previous episode.

We assume that in the beginning of an episode, the disk has already been 
formed and neglect the stage for the formation of the disk.
It is expected that during the formation stage, the
accretion rate must be very insignificant so that the influence on black hole 
spin evolution is negligible. Also, we implicitly assumed that the time lag between 
the successive episodes is long enough to let the disk be consumed on the accretion 
time scale before the next episode starts. This can be verified
by the fact that the observed ``duty cycle'' is generally much less than unity (e.g.,
\citealt{Shankar2009, Li2012}). Duty cycle measures the fraction of active black
holes to total black holes, and equivalently indicates the time fraction of active phases
over total lifetime (active and inactive) for a single
black hole (\citealt{Wang2006}).

In following calculations, unless stated otherwise, we set $\alpha=0.1$, $h=5\times10^{-3}$,
$ R_{\rm d}=2.7\times10^3M_8^{-2/3}\eta_{-1}^{2/3}R_{\rm g}$,
$M_{\rm h}=10^8M_\odot$, and $a=0.5$ as the fiducial values.
It is worth mentioning that the choice of $R_{\rm d}$
dependent on black hole mass and the radiative efficiency aims to
let the dimensionless accretion rate $\dot M_{\rm h}/\dot M_{\rm Edd}$ at about 0.25,
the typical value observed in luminous AGN surveys (\citealt{Kollmeier2006, Shen2008, Li2011}).
We also set an upper limit of mass accretion rate by $\dot M_{\rm h}\leq\dot M_{\rm Edd}$ 
considering that beyond the Eddington limit, radiative feedback may be important to regulate
the accretion rate at around the Eddington limit.

\subsection{Alignments and Anti-Alignments}
In Figure \ref{fig_angle}, we demonstrate evolution of the inclination angle between
the disk and the black hole for different lifetimes of gas fueling:
$\Delta T_{\rm f} = (0, 1, 2, 4, 10)\times 10^5$yr.
The initial value of $\theta$ is set to be $\theta_0=\pi/6$ and
$\theta_0=5\pi/6$ in the left and right panels, respectively.
As can be seen, for the case of $\theta_0=\pi/6$, the inclination
angle rapidly decreases on a timescale of $10^5$yr, which is
the typical estimate in previous numerous literature (
see, e.g., \citealt{Perego2009, Li2013a} and references therein).
As the lifetime of gas fueling $\Delta T_{\rm f}$ increases,
the decay rate of $\theta$ is slightly reduced. This is because
during the gas fueling, the disk maintains its orientation and
only the black hole is driven to align to the disk. While after
the gas fueling is quenched, both the disk and the hole change their
orientations towards the direction of the total angular momentum
of the system. This can be further verified from Equations~(\ref{eqn_angle_p1})
and (\ref{eqn_angle_p2}).
Interestingly, the behavior of $\theta$ for $\Delta T_{\rm f}=10^6$yr,
namely, the disk is continuously fed throughout the time range of calculations,
is exactly identical to the analytic solution found in \cite{Scheuer1996} and
\cite{Martin2007}.

The evolution of $\theta$ is much more complicated for the case of $\theta_0=5\pi/6$ as shown
in the right panel of Figure \ref{fig_angle}. With no gas fueling ($\Delta T_{\rm f}=0$),
the inclination angle rises up to $\pi$ all the way.
Indeed, the instantaneous behavior of $\theta$ is governed by
the ratio $J_{\rm d}/J_{\rm h}$ in Equation~(\ref{eqn_angle_p2}):
for $J_{\rm d}/J_{\rm h} > -\cos\theta$, the inclination $\theta$ decreases, while
for $J_{\rm d}/J_{\rm h} < -\cos\theta$, the inclination $\theta$ increases
(see also \citealt{Li2013a}).
As $\Delta T_{\rm f}$ increases, the final configuration of the system
transits from anti-alignment ($\theta=\pi$) to co-alignment ($\theta=0$).
We note there is a steep decreasing/increasing trend of $\theta$
just after the gas fueling is quenched, ascribed to the rapid
reduce of the angular momentum of the disk and hence enhancement of mass accretion rate
due to the viscous dissipation arising from alignments. See below for
detailed explanations.

Figure \ref{fig_rate} shows the corresponding mass accretion rates.
For the sake of comparison, the asymptotic evolution of accretion rate by
Equation (\ref{eqn_mdot_asym}) is superposed in the left panel.
If there is no gas fueling, as the inclination angle
tends to zero, the angular momentum of the disk approach constant.
As a result, the mass accretion rate evolves asymptotically following
Equation (\ref{eqn_mdot_asym}). In the right panel,
the initial value of $\theta$ is $\theta_0=5\pi/6$.
Instead of a gradual decay, there exists a notable peak of accretion rate, which
even reaches the Eddington limit $\dot M_{\rm Edd}$ in some cases.
The similar behavior of an enhancement of the mass accretion rate also appears
in  the previous numerical simulations on warped accretion disks
(see, e.g., \citealt{Lodato2006} and \citealt{Nixon2012}).
This can be easily understood in terms of Equations (\ref{eqn_jds}),
(\ref{eqn_rd}), and (\ref{eqn_mdot1}): strong dissipation arising from
alignments leads to an intense reduce of the magnitude of $J_{\rm d}$;
the effective radius $R_{\rm d}$ accordingly shrinks to maintain a
Keplerian rotating disk. As a consequence, the accretion time scale
$t_{\rm acc}$ is shortened and mass accretion rate $\dot M_{\rm h}$ is significantly
enhanced. In reality, emergence of the enhancement of accretion rate depends on competition
between the viscous-dissipation driven deduce of $J_{\rm d}$ (due to warp alignments)
and mass-accretion driven deduce of $M_{\rm d}$. This is the reason why there is
no accretion rate enhancement for some curves in Figure \ref{fig_rate}.

By assuming the angular momentum conservation of the system,
\cite{King2005} derived a condition for occurrence
of anti-alignments: the initial angle between the disk
and the hole satisfies $\cos\theta_0<-J_{\rm d}/2J_{\rm h}$.
This equation is intensively used in subsequent studies
on cosmological evolution of SMBH spins (e.g.,
\citealt{Volonteri2007, King2008, Lagos2009,
Fanidakis2011, Dotti2013, Dubois2014, Sesana2014}).
%Unfortunately, the definition of $J_{\rm d}$ is somewhat vague in
%\cite{King2005}.
%Subsequent studies mainly introduce two interpretations for
%$J_{\rm d}$: the first is the angular momentum of the disk within some radius
%(e.g., the radius where the self-gravity of the disk is dominated so
%that the disk is truncated; \citealt{Dotti2013}) and the other is the total angular
%momentum passing through the warp radius during the accretion lifetime
%(e.g., \citealt{Volonteri2007, King2008}).
%It is trivial to verify that the resulting magnitudes of $J_{\rm d}$ from these
%two approaches are quite different since the warp radius and the self-gravitational radius
%generally differs by an order of two magnitudes (see e.g., \citealt{Perego2009}).
%Moreover, both these two interpretations have their own shortcomings.
%As argued by \cite{Volonteri2007}, the first one is a static condition and does not
%include the dependence on the properties of the black hole or of the accretion disk.
%Whereas the second one is even more worse. Apparently it does not satisfy the
%assumption of the conservation of the angular momentum, the basis of \cite{King2005}'s condition,
%because the total angular momentum is continuously flowing through
%the warp radius with mass accretion and thus is no longer conserved.
We relax the assumption of the conservation of
angular momentum and let the accretion disk steadily fueled for
a lifetime of $\Delta T_{\rm f}$.
Since the total angular momentum of the system is not conserved,
it is inappropriate to directly use the \cite{King2005}'s condition.
In the right panel of Figure \ref{fig_angle}, the initial ratios $J_{\rm d}/J_{\rm h}=0.40$
and $\cos\theta_0=-0.87$, therefore the condition $\cos\theta_0<-J_{\rm d}/2J_{\rm h}$
is well established. However, for $\Delta T_{\rm f}\gtrsim2\times10^5$yr,
the system ends up an alignment instead of an anti-alignment.

Nevertheless, the ratio $J_{\rm d}/J_{\rm h}$ remains an important indicator
of alignments or anti-alignments (if there is no gas fueling).
Small ratios mean that the angular momentum of black holes dominates over
that of disks, therefore it is comparatively easier to drive the disks
towards alignments or anti-alignments. Conversely, if the angular momentum
of the disks is dominated, the holes are ``lighter'' to change their orientations.
This is somehow similar to the cases that the disks are continuously fueled:
the holes tend to align towards the disks
(see also the discussions of \citealt{King2005} and \citealt{Lodato2006}).

\begin{figure}[t!]
\centering
\includegraphics[angle=-90.0, width=0.48\textwidth]{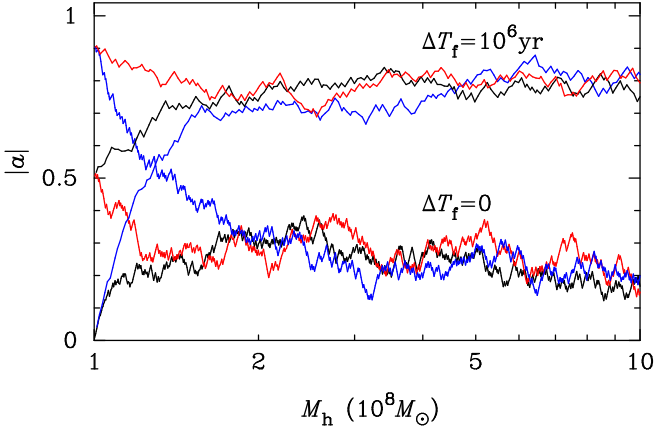}
\caption{Evolution of spin magnitude with mass growth
during a series of accretion events. The initial spin is
$a=0.9$, $0.5$, and $0.0$ for red, blue, and black lines, respectively.
The gas fueling lifetime is $\Delta T_{\rm f}=10^6$yr
for upper three lines, and there is no gas fueling for lower three lines.}
\label{fig_spin_time}
\end{figure}

\subsection{Spin Evolution}

\subsubsection{A Single Accretion Event}

In Figures \ref{fig_mass} and \ref{fig_spin}, we show evolution of
the black hole's properties, i.e., mass and spin, respectively,
for initial inclination angles $\theta_0=\pi/6$ and $\theta_0=5\pi/6$
with a set of gas fueling lifetimes $\Delta T_{\rm f}$.
Compared with the case of $\theta_0=\pi/6$, we can find that
the black hole mass growth for  $\theta_0=5\pi/6$ is much more
efficient. This is because the Eddington limit is
larger with lower radiative efficiency due to retrograde accretion for $\theta>\pi/2$
and also because the accretion rate is enhanced as plotted in
Figure \ref{fig_rate}. The spin evolution in Figure \ref{fig_spin}
clearly illustrates the important influences of gas fueling.
As the lifetime of gas fueling increases, the black hole is gradually
driven to align with the disk regardless of the initial inclination.
As a result, in right panel of Figure \ref{fig_spin}, there exists a trend
for $\Delta T_{\rm f}\gtrsim2\times10^5$yr
that spin first declines due to retrograde accretion and then
increases when the inclination angle transits to $\theta<\pi/2$.
In addition, it can be found that the spin changes are relatively more
significant for black holes with $\Delta T_{\rm f}<2\times10^5$yr,
because these holes undergo retrograde accretion throughout
the episode, which carries larger angular momentum compared
with the prograde accretion (see Equations (\ref{eqn_lms1}) and (\ref{eqn_lms2})).

\subsubsection{A Series of Accretion Events}
The spin of black holes will reach an equilibrium value after a
series of accretion events, depending on the fraction of prograde
and retrograde accretion (\citealt{King2008, Sesana2014}).
Evidently, this fraction is intimately relevant to the degrees of
anisotropy in the orientations of accretion disks with respect to
the holes. If all the accretion events occur with the initial
inclination angles $\theta_0<\pi/2$, the black holes
accrete in a coherent fashion and their spin will be spun up
all the way to the maximum ($a=1$). On the other hand, the inclination
angles $\theta_0$ may distribute randomly over the whole space $(0\sim\pi)$
(e.g., \citealt{King2008, Wang2009, Hopkins2012, Li2012}).
For simplicity, we only focus on the random distribution
of $\theta_0$ in proportional to $\sin\theta_0$ over $(0\sim\pi)$,
so that both alignments and anti-alignments
can occurs in our calculations. We first generate a random inclination
angle and then apply the prescription described above to track down
the mass growth and spin evolution of black holes. The calculation ends when
the accretion rate decreases to one tenth of its initial value,
i.e., $0.025\dot M_{\rm Edd}$.
%Such procedure is repeatedly run for a
%sufficiently large times to obtain a desired mass range of black holes.

In Figure \ref{fig_spin_time}, we illustrate how black hole spin evolves
along their mass growth during a series of accretion events.
We stress that we adopt the effective radius of accretion disks dependent
on black hole mass as $R_{\rm d}=2.7\times10^3M_8^{-2/3}\eta_{-1}^{2/3}R_{\rm g}$
in order to guarantee the initial dimensionless accretion rate at $0.25$ in
Equation (\ref{eqn_mdot_edd}).
Once there is no gas fueling, irrespective of the initial spin values,
black holes lose the memory of their initial status and
approach a relatively low spin roughly after doubling their mass.
Then as mass growth, there is a weak trend that their spin gradually
falls off. Let's give an intuitive explanation for such weakly negative dependence
of spin on mass through the ratio $J_{\rm d}/J_{\rm h}=1$.
Since we adopt $R_{\rm d}\propto M_{\rm h}^{-2/3}$, Equation (\ref{eqn_ratio})
yields $a\propto M_{\rm h}^{-1/3}$. Of course, the real situation is
much more complicated, but the negative dependence is retained (see also below
for further discussions).

Once the disks are fueled for a lifetime longer that the typical alignment timescale (e.g.,
$\Delta T_{\rm f}=10^6$yr), as expected, the holes are always aligned to the disks and
are rapidly spun up to fast rotating. However, we note that the final spin
never reaches the maximal value $a=1$ because during the alignments, there
exists a period with retrograde accretion that spins down the holes.
This indicates that even though the final configuration in an accretion event
is alignment instead of anti-alignment, the spin decrease during the time with
retrograde accretion still contributes to the final spin status and therefore should
be appropriately included.

\begin{figure*}[t!]
\centering
\includegraphics[angle=-90.0, width=0.45\textwidth]{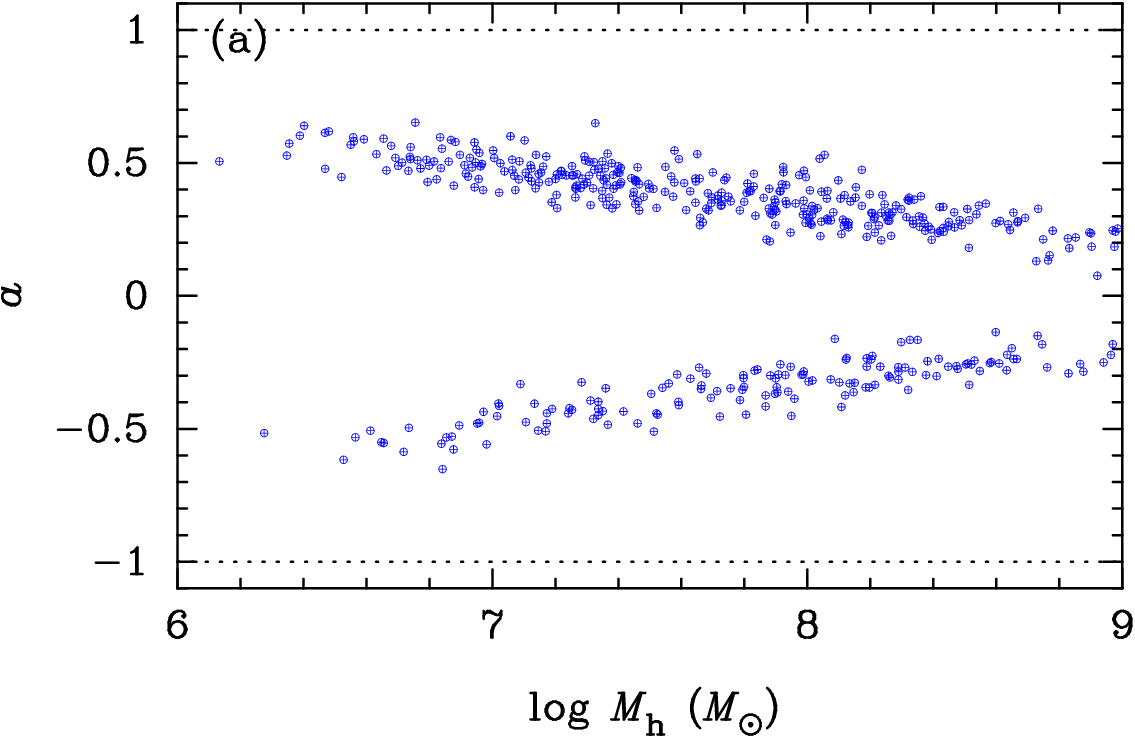}~~~~
\includegraphics[angle=-90.0, width=0.45\textwidth]{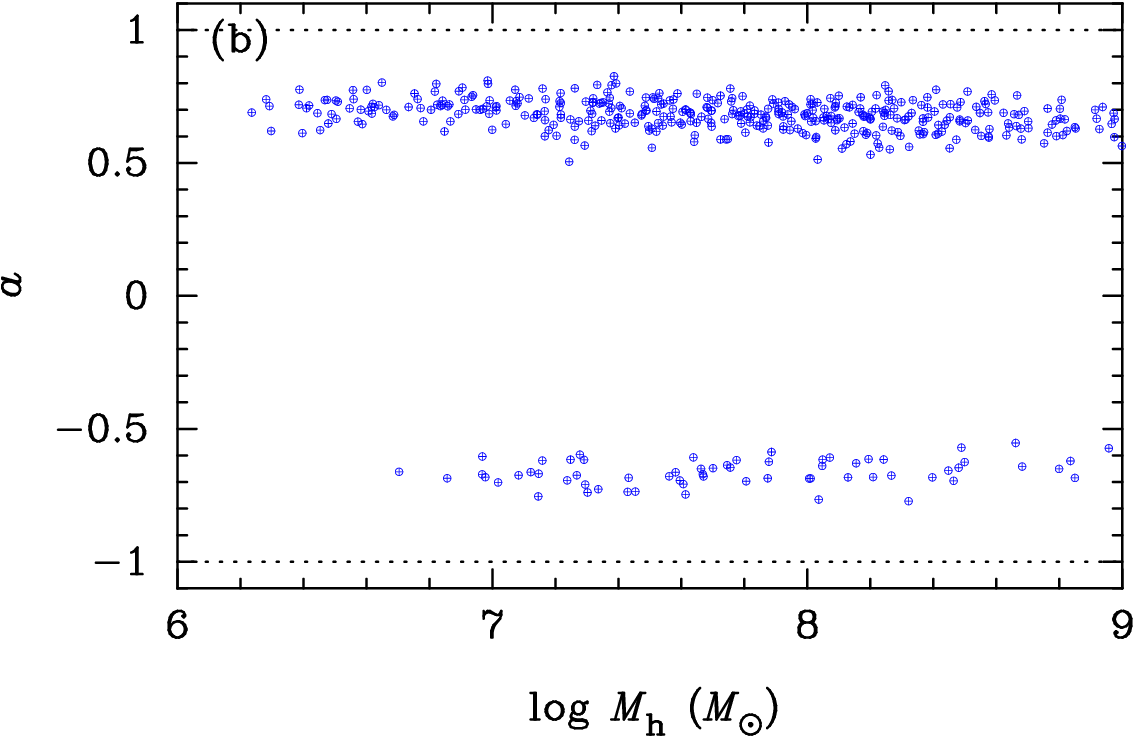}\\\vspace*{0.4cm}
\includegraphics[angle=-90.0, width=0.45\textwidth]{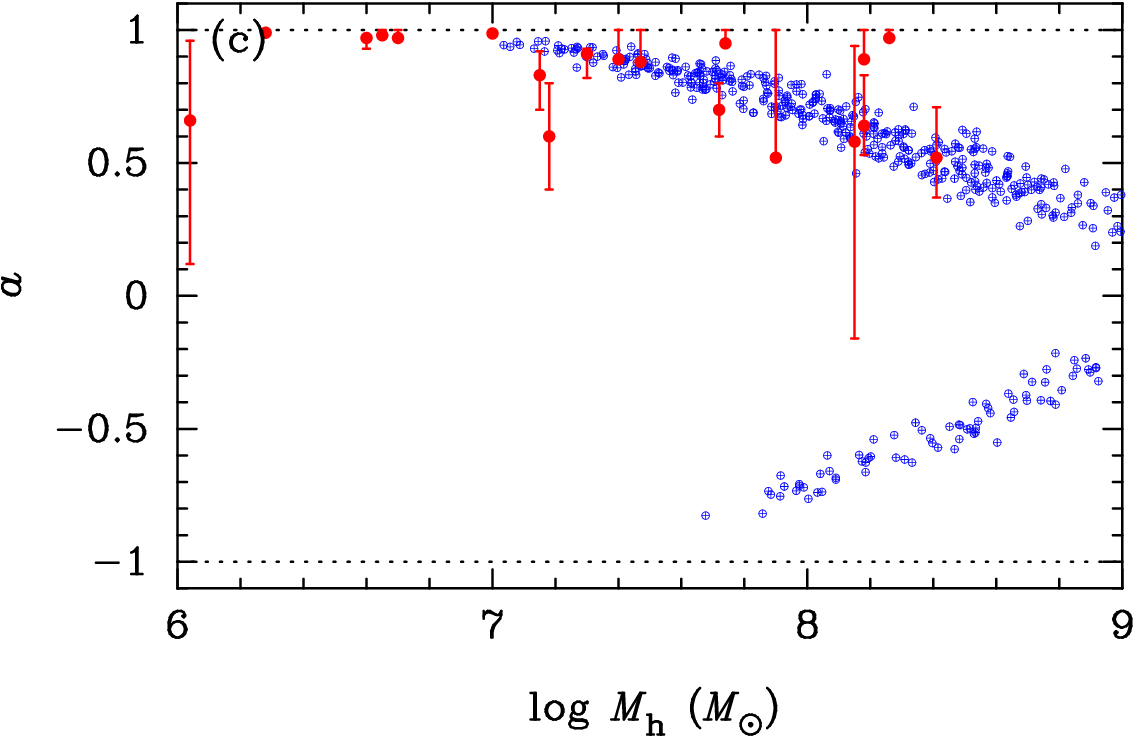}~~~
\includegraphics[angle=-90.0, width=0.45\textwidth]{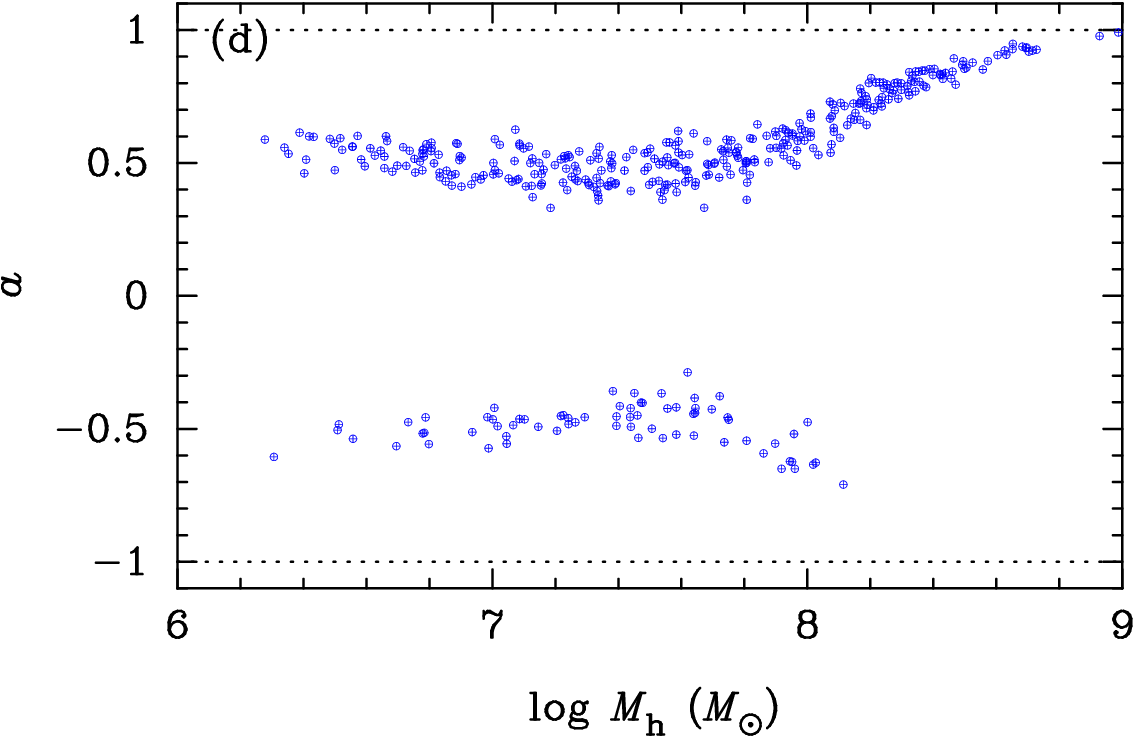}
\caption{Spin distribution with black hole mass from Monte-Carlo simulations
for different gas fueling lifetime $\Delta T_{\rm f}$:  (a) $0$, (b) $10^5$yr,
(c) $10^5M_8^{-1}$yr, and (d) $10^5M_8^{1.5}$yr.
Negative spin means that black holes are undergoing retrograde accretion.
Data points with error bars in panel (c) are the measured spin through
broad Iron K$\alpha$ line compiled by \cite{Brenneman2013} and \cite{Reynolds2013}
(see also Table 2 of \citealt{Sesana2014}). Horizontal dotted lines represents
black hole spin $a=1$ and $-1$, respectively.}
\label{fig_spin_dist}
\end{figure*}
\begin{deluxetable*}{ccccc}
\centering
\tablecolumns{5}
\tablewidth{0pt}
\tablecaption{Episodic Lifetimes of AGNs Constrained from Observations in the Literature.}
\tablehead{
  \colhead{Object}
& \colhead{Redshift}
& \colhead{Episodic Lifetime (Myr)}
& \colhead{Telescope}
& \colhead{Reference}
}
\startdata
\multicolumn{5}{c}{Proximity Effect}\\\hline
Q 0302-003         &  3.285      &  $10-30$  &  VLT                      &  1, 2\\
HE 2347-4342       &  2.885      &  $\sim25$     &  VLT                      &  3\\
KP 76              &  2.466      &  $\gtrsim25$        &  Keck, HIRES              &  4\\
KP 77              &  2.535      &  $16-33$  &  Keck, HIRES              &  4\\
HS 1700+6416       &  2.748      &  $\gtrsim20$        &  HST/COS                  &  5\\
Quasar Pair Sample &  $\sim2.2$  &  $\sim 1$     &  Keck, Lick, NOAO, SDSS   &  6\\
\cutinhead{Fluorescent Ly$\alpha$ emission}
HS 1549+1919       &  $\sim2.84$ &  $\gtrsim1.3$       &  Keck                     &  7\\
QSO 0420-388       &  $\sim3.1$  &  $\gtrsim60$        &  VLT                      &  8\\
Quasar Sample      &  $\sim2.7$  &  $1-20$       &  Keck                     &  9
\enddata
\tablerefs{(1) \cite{Worseck2006}; (2) \cite{Jakobsen2003}; (3) \cite{Worseck2007};
(4) \cite{Goncalves2008}; (5) \cite{McQuinn2014}; (6) \cite{Kirkman2008};
(7) \cite{Adelberger2006}; (8) \cite{Cantalupo2007}; (9) \cite{Trainor2013}.
}
\label{tab_lifetime}
\end{deluxetable*}

\section{Implications for Spin Distribution And Episodic Lifetimes of AGNs}
\subsection{Monte-Carlo Simulations}
We perform Monte-Carlo simulations to explore the spin distribution
with the above described prescription.  The free parameters $\alpha$,
$h$, and $R_{\rm d}$ are set by their fiducial values as in Section~\ref{sec_results}.
Initially, we assign to a black hole a mass drawn from a uniform distribution
in logarithm over $(10^6\sim10^8)M_\odot$ and
a spin drawn uniformly over $(0\sim1)$.
For this newly generated black hole, we first let it grow with 100 accretion
episodes, to remove the memory of its initial conditions.
Afterwards, this hole continues to accrete for a series of episodes.
We randomly assign the number of episodes between $(0\sim900)$ so that in average
the hole undergoes 500 times activities.
We then record the final mass and spin information of the hole.
In each accretion episode, the inclination $\theta_0$ between the hole and its accretion disk
follows a uniform distribution of $\cos\theta_0$ over $(-1\sim1)$.
An individual event is regarded to be terminated once the accretion rate
decreases below a tenth of its initial value.
We repeat such procedure 500 times and are finally left with a sample of 500 black holes.

In Figure \ref{fig_spin_dist}, we plot the spin distribution with black hole mass for
different gas fueling lifetime $\Delta T_{\rm f}=0$, $10^5$yr,
$10^5M_8^{-1}$yr, and $10^5M_8^{1.5}$yr in panels (a)-(d), respectively.
In panel (a), there is no gas fueling and the disks will be consumed
within an accretion timescale $t_{\rm acc}\sim8.8\times10^5$yr
(see Equation (\ref{eqn_tacc})). As the ratio $J_{\rm d}/J_{\rm h}$
decreases with black hole mass, the accretion disks of massive black holes
are easier change their orientations and thus anti-alignments
become more probable. This results in a negative
correlation between spin and mass. A similar trend
was also found by \cite[corresponding to $F=0$ in their model]{Dotti2013}.
When we switch on the gas fueling for a time of $\Delta T_{\rm f}=10^5$yr
in panel (b), black holes are always driven to align towards the
disks during this period. Therefore, the fraction of retrograde
accretion is significantly reduced and all the black holes carry
spin $|a|\gtrsim0.5$.

We also choose a mass-dependent lifetime of gas fueling for the sake of
comparison with observations.
In panel (c), the gas fueling lifetime is $\Delta T_{\rm f}=10^5M_8^{-1}$yr,
indicating that the disks surrounding $10^7M_\odot$ black holes will be
fueled for $10^6$yr, far longer than the alignment timescale.
Accordingly, smaller black holes are overall spun up to maximum rotating.
There is a strong anti-correlation between spin and mass.
On the other hand, in panel (d), we reverse the mass dependence of fueling lifetime
to $\Delta T_{\rm f}=10^5M_8^{1.5}$. As expected, massive black holes at
$\sim10^9M_\odot$ have maximum spin $(a=1)$ whereas lower massive black
holes in this case are not affected (since $T_{\rm f}$ is much less that
the accretion timescale and can be neglected) and just follow the
distribution in panel (a).

\begin{figure}[t!]
\centering
\includegraphics[angle=-90.0, width=0.45\textwidth]{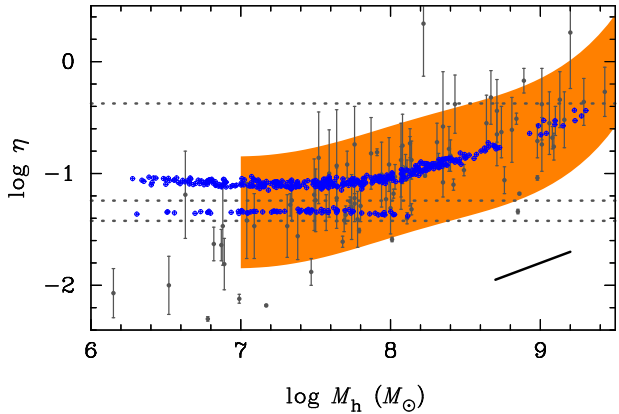}
\caption{Radiative efficiency with black hole mass for gas fueling lifetime
$\Delta T_{\rm f}=10^5M_8^{1.5}$yr. Negative spin
means that the black holes are undergoing retrograde accretion.
Gray points with errorbars
are the radiative efficiency of local quasars estimated by \cite{Davis2011} and shaded area
represents the radiative efficiency at $z\sim1$ estimated by \cite{Li2012}.
Horizontal dotted lines represents $\eta=0.42$, $0.057$, and $0.038$ for black hole spin
$a=1$, $0$, and $-1$, respectively.
To guide the eye, a solid line with a slope of 0.5 is plotted
in the bottom right corner. }
\label{fig_eta}
\end{figure}

\subsection{Comparison with Observations}
We now compare our calculated spin distributions with observations.
Notwithstanding measuring spin of massive black holes is yet challenging so far,
there are several developing techniques shedding useful light
(e.g., \citealt{Wang2009, Czerny2011, Davis2011, Li2012, Brenneman2013, Done2013,
Reynolds2013, Wu2013}). Among these techniques,  we focus on
the direct measurements using the broad Iron K$\alpha$ line fitting
(\citealt{Brenneman2013, Reynolds2013} and references therein)
and the indirect estimates using the radiative efficiency
of thin accretion disks as a surrogate of the black hole spin
(see Equation (\ref{eqn_eta}); \citealt{Davis2011, Li2012}).
Both these two methods had been applied to a sample or a population of AGNs,
allowing us to perform straightforward comparison.
Unfortunately, each method bears a large source of uncertainties that are not well understood
and the obtained results seem incompatible at this stage.
Therefore, we perform comparison separately as follows.

In panel (c) of Figure \ref{fig_spin_dist}, we superpose the spin of 19
sources measured from Iron K$\alpha$ lines, most of which are Seyfert I or
narrow line Seyfert I AGNs residing in spirals or lenticular galaxies
(see \citealt{Sesana2014}). \cite{Reynolds2013} proposed
that in this sample low massive black holes ($\sim10^7M_\odot$)
are spinning fast while black holes with mass $>10^8M_\odot$ may
have moderate spin. Our calculations show that a lifetime of gas fueling
as $\Delta T_{\rm f}=10^5M_8^{-1}$ can generally reproduce such observed
spin distribution. This means  low massive black holes require longer gas fueling to
reach alignments and to accrete coherently.

On the other hand, \cite{Davis2011} reconstructed the spectral energy distributions
for a sample of 80 Palomar-Green quasars and estimated their individual radiative efficiency
of accretion flows based on the thin disk model. They found a tight correlation
between the efficiency with black hole mass approximately as
$\eta\propto M_{\rm h}^{1/2}$ (but see also \citealt{Raimundo2012}), plausibly
implying a raise in the black hole spin with mass.
Meanwhile, \cite{Li2012} applied the \citeauthor{Soltan1982}'s argument (\citealt{Soltan1982})
to solve the continuity equation of SMBH demography in active and quiescent galaxies
(\citealt{Li2011}) and quantified the radiative efficiency
with redshift and black hole mass. Their results suggested that the efficiency
undergoes cosmological evolution and generally there is an increasing trend of the efficiency
with black hole mass at redshift $z\gtrsim1$ (see also \citealt{Cao2008, Volonteri2013, Ueda2014}).
Figure \ref{fig_eta} compares the radiative efficiency
from our calculations by adopting the gas fueling lifetime $\Delta T_{\rm f}=10^5M_8^{1.5}$yr
(as in panel (d) of Figure \ref{fig_spin_dist}) with these from observational constraints of
\cite{Davis2011} and \citet[at $z\sim1$]{Li2012}. Here, the radiative efficiency
is calculated from black hole spin as Equation (\ref{eqn_eta}) by assuming that the thin disk model applies.
As stated above, in the context of chaotic accretion without gas fueling,
high massive black holes ($\sim10^9M_\odot$) have low spin because of
net spin-down by retrograde accretion with anti-alignments.
Therefore, to maintain the observed high radiative efficiency (and high spin),
sufficient prolonged gas fueling is in need for these holes.

Longer gas fueling means that there is plenty of gas reservoir progressively channeled
into the accretion disks and the black holes are accordingly long-lived.
There are several techniques developed for measuring or estimating episodic lifetimes
of AGNs, see \cite{Martini2004} for a review.
In Table~\ref{tab_lifetime}, we summarize the observational constraints on
episodic lifetimes of AGNs through the proximity effect and the  fluorescent Ly$\alpha$ emission.
Note that both these constraints are presently only realizable for high-redshift, luminous quasars, indicating that
there reside high massive black holes (plausibly $>10^9M_\odot$).
It seems that the observed AGN episodic lifetimes are generally $\sim10^7$yr but with large uncertainties.
Such long lifetimes strongly imply that high massive black holes are rapidly spinning
at high redshift, consistent with previous studies probed through the radiative efficiency
(e.g., \citealt{Wang2009, Li2012, Trakhtenbrot2014}). Our calculations in the present work
further show that if the correlation between the efficiency and the black hole mass is reliable,
low massive black hole must accrete in more chaotic way and a gas fueling lifetime as
$\Delta T_{\rm f}\sim 10^5M_8^{1.5}$ can roughly reproduce the observed slope
$\eta\propto M_{\rm h}^{1/2}$. This gives rise to an episodic lifetime of
$>3\times10^6$ yr for SMBHs with $M_{\rm h}>10^9M_\odot$, coincident with the observations.

%To summarize, the gas fueling lifetime, which in effect controls the observed episodic
%lifetime of AGNs,  plays an important role not only on the mass growth of SMBHs, but also
%on their spin evolution in terms of the Bardeen-Petterson effect.

\section{Discussions and Conclusions}
\begin{figure}[t!]
\centering
\includegraphics[angle=-90.0, width=0.45\textwidth]{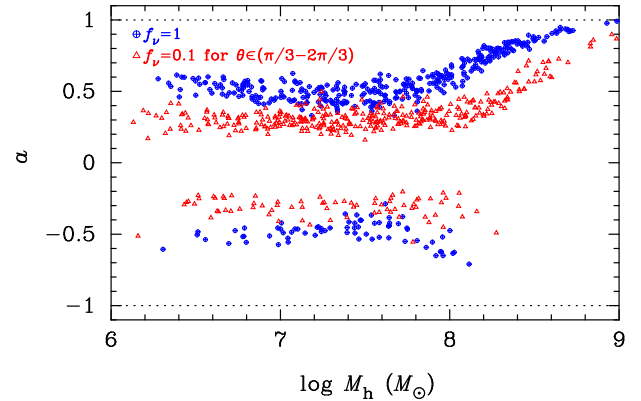}
\caption{Influence of $f_\nu$ on spin distribution. Blue points represent the case
$f_\nu=1$ for all the $\theta$s and red points represent the case $f_\nu=0.1$ for $\theta\in(\pi/3-2\pi/3)$.
The gas fueling lifetime is $\Delta T_{\rm f}=10^5M_8^{1.5}$yr.}
\label{fig_fnu}
\end{figure}

The main differences of the present work compared with the previous studies
(e.g., \citealt{Perego2009, Dotti2013}) lie at:
1) we include the influences of continuous gas fueling on the alignments
between black holes and their accretion disks; and 2) we consider the finite-size
accretion disks (see also \citealt{King2008}).
There are several simplifications in our calculations  meriting further explanations.
First, we neglect the detailed structures and properties of accretion disks and employ
an approximate global model to simulate  the evolution of disks under
alignments and accretion (\citealt{Kumar2008}). We are thus unable to handle
the angular momentum transportation throughout the disks.
This may be important when the disks suffer large warps  ($\theta\sim\pi/2$)
in which non-linear effects cause the disk structures around the warp radius to change
dramatically and the simple global approximate plausibly becomes invalid. However, since
the spin parameter invokes the whole accretion history, we expect that our results
are insensitive to this effect. Secondly, the main free parameters in our 
calculations are the disk's aspect ratio
$h$ and the effective radius $R_{\rm d}$. In terms of the mass accretion rate,
$h$ and $R_{\rm d}$ is degenerated along lines of constant $h^2R_{\rm d}^{-1}$.
As mentioned above, the standard accretion disk model shows
that $h$ is almost insensitive to radius and only weakly depends on black hole mass
and mass accretion rate (\citealt{Natarajan1998, King2008}). We therefore fix $h$ to 
the intermediate value between $10^{-3}\sim10^{-2}$.  The effective radius is chosen by 
a specified mass accretion rate in Equation (\ref{eqn_mdot_edd}), which is roughly 
at $\sim0.25\dot M_{\rm Edd}$ from AGN surveys. We recalculate the spin distribution 
for $h$ between $10^{-2}$ and $10^{-3}$ and find the results are qualitatively unchanged.
Thirdly, the Lense-Thirring torque $K_2$ calculated by Equation (\ref{eqn_K2_mdot}) 
is also a well approximate for small warps (\citealt{Lodato2006}). When disks are 
strongly warped, on one hand, the relation $\dot M\approx3\pi\nu_1\Sigma$ no 
longer applies, in particular in the region around the warp radius; 
on the other hand, the vertical shear viscosity $v_2$ deviates from the
relation in Equation (\ref{eqn_vis}) and turns to be decreasing as warp amplitude increases 
(\citealt{Ogilvie1999, Lodato2010}). To see how our results depend on $f_\nu$, we 
artificially set an extreme value of $f_\nu=0.1$ for large inclination angle 
$\theta\in(\pi/3-2\pi/3)$ and show the comparison with $f_\nu=1$ in Figure~\ref{fig_fnu}. 
The choice of a such range is based on the study by \cite{Perego2009}
who found that the non-linear term becomes important for angular momentum transportation
when the inclination angle $>\pi/3$. Black hole spin magnitude is systematically smaller
because of the reduced torque $K_2$ for large warps.
However, we can find that the changes are moderate with such an extreme $f_\nu$ value and the overall
spin distribution with mass is preserved.

For simplicity, we also assume that disk fueling is chaotic so that the initial inclination angle
$\theta_0$  between the disks and black holes distributes randomly over $(0\sim\pi)$.
\cite{Dotti2013} proposed a comprehensive modeling on the anisotropy of $\theta_0$
using a parameter $F$ to describe the fraction of accretion events
with $\theta_0>\pi/2$. Therefore,  $F=0.5$ stands for fully chaotic fueling and
$F=0$ stands for fully coherent accretion. \cite{Sesana2014} further extended this prescription
and connected the anisotropy parameter $F$ with the dynamic properties of the host galaxies.
They postulated that the ratio of the rotation velocity to the velocity dispersion of the galaxies
mirrors the degree of anisotropy of the gas component channeled into the nuclear regions. They argued
that the rotation velocity measures the bulk rotation of the galaxies while
the velocity dispersion measures how chaotic the galaxies are. This seems
in line with the recent finding of \cite{Ho2014} that the virial factor in reverberation mapping,
which encodes the geometric and kinetic properties of the broad-line regions of AGNs,
is related with the large-scale morphology of the host galaxy bulges.
However, numerical simulations showed that the angular momentum of gas inflowing
into the nuclear regions ($\sim$pc) will lose its memory of the initial direction on the
the larger scales ($\sim$ kpc; \citealt{Barnes1996, Hopkins2012}).
This is well supported by the observations that AGNs are misaligned with the
host galaxies (e.g., \citealt{Kinney2000, Gallimore2006, Shen2010, Lagos2011}).
These conflicts reflect our poor understanding on the detailed
processes in black hole fueling from the galaxy scale to the central disk scale.
We await future observations to justify these conflicts and we content with our
present assumption.

We consider only thin accretion disks with the disks' aspect ratio $H/R\ll\alpha$
to guarantee that the viscosity is  strong enough to validate the Bardeen-Petterson
effect (\citealt{Papaloizou1995}). Therefore, our approach only applies to the accretion disks
with dimensionless accretion rate $\dot M_{\rm h}/\dot M_{\rm Edd}$ at $10^{-2}\sim1$, which are
believed to correspond to the thin disk regime. Beyond this range, accretion disks transit to be thick
and the Bardeen-Petterson effect may be weak or even disappear, as confirmed by recent
magnetohydrodynamics (MHD) numerical simulations (e.g., \citealt{Fragile2009, Zhuravlev2014}).
However,  the heuristic MHD simulations by \cite{McKinney2013} revealed a new
``magneto-spin alignment'' mechanism that works in thick disks with strong magnetic.
Unlike the Bardeen-Petterson effect, this mechanism arises from the magnetic torque
of the black hole's magnetosphere, which is already aligned with the black hole
spin axis due to the frame-dragging forces. It is unknown yet how the spin evolves for
thick warped disks and more investigations are highly worthwhile.

In conclusion, we study the alignments of black holes and their accretion disks by taking into
account the finite sizes of disks and the continuous gas fueling.
%The finite sizes of disks mean that their angular momentum
%is also finite and thus is subject to change in its direction during the alignments, unlike the previous
% studies that assumed that disks are infinitely extended (e.g., \citealt{Martin2007, Perego2009}).
%The gas fueling with a lifetime means that the disks are replenished so that they maintain preferable
%profiles during the lifetime. This has an important influence on the alignments since during the period with
%gas fueling, black holes are always aligned toward the disks.
%We employ an approximate global prescription following
%\cite{Kumar2008} to calculate the mass accretion rate without digging into the detailed
%structures of accretion disks.
%As a consequence, we can explore the
%spin evolution during the alignment processes. In particular,  when the black hole-disk system
%approach alignments from an initial inclination angle $\theta>\pi/2$, accretion transits from
%retrograde to prograde fashion (see also \citealt{Li2013a}). The spin-down by retrograde
%accretion should be appropriately included to calculate the final net black hole spin.
Our results show that, with fiducial values for the free parameters, the lifetime of gas fueling
is crucial to the alignments/anti-alignments and therefore to the spin evolution.
By applying our prescription to a series of accretion activities and assuming that
the disk orientations are chaotically distributed over episodes, we compared our
calculated spin distribution with the spin measurements through Iron K$\alpha$ line fitting
and through the radiative efficiency and made attempt to place constraints on the lifetime of gas fueling.
Unfortunately, the spin measurements through these two techniques at this stage seems incompatible to
allow us draw a firm conclusion. We find that generally a lifetime of gas fueling as
$\Delta T_{\rm f}=10^5M_8^{-1}$yr can reproduce the spin distribution reported by the Iron
K$\alpha$ line fitting, whereas a lifetime as $\Delta T_{\rm f}=10^5M_8^{1.5}$yr can reproduce
these constraint through the radiative efficiency. However, the later case seems consistent with
the observed episodic lifetime as long as $\sim10^{7}$yr for very luminous AGNs at high redshift, which plausibly
harbor $>10^9M_\odot$ SMBHs.
Since the lifetimes of gas fueling are linked to the episodic lifetimes of AGNs, we proposed that
the episodic lifetimes should be regarded as a new ingredient for the semi-analytic models
of SMBH growth and spin evolution.

%===================================================================================
\acknowledgements{We thank the referee's suggestions that significantly improve the manuscript.
LYR thanks Monica Colpi, Marta Volonteri and Enrico Barausse for useful discussions
on black hole spin evolution, and the Institut d'Astrophysique de Paris where this work was completed.
The research is supported by NSFC-11133006,
11173023, 11233003, and 11303026, the China-Israel NSFC-ISF 1136114034,
and the Strategic Priority Research Program -
The Emergence of Cosmological Structures of the Chinese Academy of Sciences, grant No. XDB09000000.}
%===================================================================================

\end{document}